\documentclass[12pt]{article}
\usepackage{amsmath,amsfonts,amssymb}
\usepackage{booktabs}
\usepackage{float}

\newcommand{\mathsym}[1]{{}}
\def\id{\protect{{1 \kern-.28em {\rm l}}}}

\def\be{\begin{eqnarray}}
\def\ee{\end{eqnarray}}

\makeatletter
\renewcommand\section{\@startsection {section}{1}{\z@}%
                                   {-3.5ex \@plus -1ex \@minus -.2ex}%
                                   {2.3ex \@plus.2ex}%
                                   {\normalfont\large\bfseries}}
\renewcommand\subsection{\@startsection{subsection}{2}{\z@}%
                                   {-3.25ex\@plus -1ex \@minus -.2ex}%
                                   {1.5ex \@plus .2ex}%
                                   {\normalfont\normalsize\bfseries}}

\makeatother

\def \foot {\footnote}
\def \bi{\bibitem}

\def \ha {{1 \over 2}}

\def \ci{\cite}
\def \N {{\mathcal N}}

\def\S{{\mathcal S} }

\def \E {{\mathcal  E}} \def \J {{\mathcal  J}}

\def \Q {{\mathcal Q}}

\def\a{\alpha}

\def\C{{\bf C}}

\def \a {\alpha}

\def\g{\gamma}
\def\ov{\over}

\def\J{{\mathcal J}}

\def\E{{\mathcal E}}

\def\l{\lambda}

\def \k {\kappa}
\def\foot{\footnote}

\def\det{\hbox{det}}
\def \ci {\cite}

\def \foot {\footnote}
\def \bi{\bibitem}
\def \ha {{1 \over 2}}

\def \fo { {1\ov 4}}

\def \l  {\lambda}

\def \N {{\mathcal N}}
\def \S {{\rm S}}

\def \N {{\mathcal N}}

\def \bi{\bibitem}
\def \la {\label}

\def \l {\lambda}
\def\foot{\footnote}

\def \sql {{\sqrt \l}}
\def \adss {$AdS_5 \times S^5~$ }
\newcommand{\rf}[1]{(\ref{#1})}
\def \ov {\over}

\def\N{{\cal N}}

\def\cc{\circ}

\def \ha{{1\ov 2}}

\def \no {\nonumber}

\def \J {\mathcal{J}}

\def \E {{\cal E}}

\def \S {{\cal S}}
\def \J {{\cal J}}

 \def \bb {\bar \beta}

\def \bi{\bibitem}
\def \la {\label}

\def \l {\lambda}
\def\foot{\footnote}

\def \sql {{\sqrt \l}}

\def \adss {$AdS_5 \times S^5$\ }

\def \ov {\over}

\def \varpi {{\rm w}}
\def \OO {{\cal O}}

\def \te {\theta}

\def \cc {{\rm f}}

\def \S  {{\rm S}}

\def \C {{\cal C}}

\def \s {\sigma}

 \def \J {{\cal J}}
 \def \S {{\cal S}}
 \def \E {{\cal E}}

  \def \zet {\zeta_3}
\def \os  {\OO({\textstyle{ 1\ov \sql}})}

 \def \sql {\sqrt{\lambda}}

\def \cc {{c }} 
\def \OO {{\cal O}}
\def \te {\textstyle}
\def \fl {\sqrt[4]{\l}}

\def \fo {{\textstyle{1 \ov4}}}
\def \rx {{\rm x}}
\def \hg {{\hat \g}}

\def \C  {{\rm C}}
\def \hC  {{\rm \hat  C}}
\def \dd  {{\rm d}}
\def \bb {{\rm b}}
\def \dDelta {2}
\def \sql {{\sqrt{\l}}}
\def \ed {\end{document}}
 \def \an {{\rm an}} \def \nan {{\rm nan}}

\newcommand{\mc}{\mathcal }

\def \oomega {w}
\def \rrho {{\varrho}}

\begin{document}


\overfullrule=0pt
\parskip=2pt
\parindent=12pt
\headheight=0in \headsep=0in \topmargin=0in \oddsidemargin=0in

\vspace{ -3cm}
\thispagestyle{empty}
\vspace{-1cm}

\rightline{Imperial-TP-AAT-2012-02}
\rightline{PI-STRINGS-242}

\begin{center}
\vspace{1cm}
{\Large\bf  
``Short'' 
 spinning strings
and structure \\
of quantum $AdS_5 \times S^5$  spectrum

\vspace{1.2cm}

  }

\vspace{.2cm}
 {M. Beccaria$^{a}$, S. Giombi$^{b}$, G. Macorini$^{c}$, 
R. Roiban$^{d}$
 and  A.A. Tseytlin$^{e,}$\footnote{Also at Lebedev  Institute, Moscow. }
}

\vskip 0.6cm

{\em 
$^{a}$Dipartimento di Matematica e Fisica ``Ennio De Giorgi'', \\ Universita' del Salento \& INFN, 
                     Via Arnesano, 73100 Lecce, Italy\\
\vskip 0.08cm
\vskip 0.08cm 
$^{b}$Perimeter Institute for Theoretical Physics, Waterloo, 
Ontario, N2L 2Y5, Canada\\
\vskip 0.08cm
\vskip 0.08cm 
$^{c}$ Niels Bohr International Academy and Discovery Center,  
		     Niels Bohr institute, \\
		     Blegdamsvej 17 DK-2100 Copenhagen, Denmark\\
\vskip 0.08cm
\vskip 0.08cm 
$^{d}$Department of Physics, The Pennsylvania  State University,\\
University Park, PA 16802 , USA\\
\vskip 0.08cm
\vskip 0.08cm 
$^{e}$Blackett Laboratory, Imperial College,
London SW7 2AZ, U.K.
 }

\vspace{.2cm}

\end{center}

\begin{abstract}
 
Using information from  the marginality conditions of vertex operators
for the $AdS_5\times S^5$ superstring, we determine the structure of the dependence of the energy of quantum string states on their conserved charges  and the string tension  $\sim\sqrt \lambda$. 
 We consider states  on the  leading Regge trajectory 
 in the flat space limit which 
  carry one or two (equal) spins  
 in $AdS_5$ or $S^5$ and an orbital momentum  in $S^5$, with Konishi multiplet states
 being particular cases.   We argue that the coefficients 
 in the energy may be found  by using a  semiclassical  expansion.
By analyzing the examples  of folded spinning strings in $AdS_5$ and $S^5$
 as well as three cases of circular two-spin strings we 
demonstrate the universality of 
 transcendental (zeta-function)  parts of few leading  coefficients. We also  show the 
consistency  with target space supersymmetry with different states 
belonging to the same multiplet having the same non-trivial part of the energy.
We suggest, in particular,  that a  rational coefficient 
(found   by Basso for the folded string using  Bethe Ansatz considerations  
and  which, in general,  is yet to be determined by 
a direct two-loop string calculation) should,  in fact,  be  universal.

\end{abstract}

\newpage
\setcounter{equation}{0} 
\setcounter{footnote}{0}
\setcounter{section}{0}
\renewcommand{\theequation}{1.\arabic{equation}}
 \setcounter{equation}{0}
\setcounter{equation}{0} \setcounter{footnote}{0}
\setcounter{section}{0}

\def \os {O(\textstyle{ {1 \ov (\sql)^2}} )}
\def \ost {O(\textstyle{ {1 \ov (\sql)^3}} )}
\def \cc {{c }} 
\def \OO {{\cal O}}
\def \te {\textstyle}
\def \fl {\sqrt[4]{\l}}

\def \ha {{{\textstyle{1 \ov2}}}}
\def \fo {{\textstyle{1 \ov4}}}
\def \rx {{\rm x}}
\def \hg {{\hat \g}}

\def \C  {{\rm C}}
\def \hC  {{\rm \hat  C}}
\def \dd  {{\rm d}}
\def \bb {{\rm b}}
\def \dDelta {2}
\def \sql {{\sqrt{\l}}}

 \def \an {{\rm an}} \def \nan {{\rm nan}}
 \def \nm {\tilde n_{11}}
 \def \tn {{\tilde n}}  \def \uni {{\rm inv}}
\def \ttn  {{\bar n}_{11}}

\section{Introduction and summary}

Recent progress in understanding the integrable system that should  be computing the spectrum of the maximally supersymmetric example of AdS/CFT duality 
  makes it important 
to further develop  a detailed matching of the Bethe ansatz predictions with quantum 
\adss  string energies extracted from the perturbative string theory. 
  While direct  near-flat-space expansion of the quantum string theory  determining 
 the  large tension ($T= { \sql \ov 2\pi}$) expansion of quantum string energies 
 with fixed quantum charges is still to be developed, 
here we shall follow the ``semiclassical'' approach  suggested in \cite{rt1}
(see also \cite{tt})  and recently applied in \cite{gssv,rt2,gva,bm}
to demonstrate the matching of  the numerical results of the TBA for the Konishi operator 
dimension 
interpolated  from weak to strong coupling 
\cite{gkv,f,ff} with the perturbative string theory prediction
for the corresponding string energy.

Our motivation is to further understand the structure of the dependence 
of the string energy on the string tension and its quantum numbers (spins)
guided by the expected  form of the string vertex operator 
marginality conditions \cite{rt1,rt2}  and recent progress 
 on the Bethe ansatz side
\cite{bas}. We shall consider several string states which belong (in the flat-space limit)
to the leading Regge trajectory and for the lowest values of the spins or the 
lowest value of the string level represent states
  in the Konishi multiplet  and 
discover  the universality of some 
leading-order coefficients in the expansion
of their  energies.

\subsection{General structure of the inverse tension expansion of 
the energy}

Let us start  with describing the general form of the dependence of the energy $E$ of a
string state on its quantum charges $Q_i$ in the large string tension expansion 
($\sql\gg 1$).\foot{Examples of these charges discussed below 
are spins $S_1,S_2$ in $AdS_5$ and spins 
$J_1,J_2,J_3$ in $S^5$.}
 As follows from the structure of $\a'$ expansion of 2d  anomalous
dimensions of the  corresponding \adss string vertex operators \cite{p,t03}, the solution
of the marginality  condition should give $E=E(Q,\sql)$ in the following 
general form
\cite{rt1,rt2}
\be 
&& E^2= 2 \sql \sum_i a_i Q_i   + \sum_{i,j} b_{ij} Q_i Q_j + \sum_i c_i Q_i\no
\\ && \ \ \ \ \ 
 + 
 {1 \ov \sql} \Big( \sum_{i,j,k} d_{ijk} Q_i Q_j Q_k  +  \sum_{i,j} e_{ij} Q_i Q_j 
 +   \sum_i f_i Q_i    \Big) + \os \ , 
\la{1} 
 \ee
where $Q_i$ are supposed to be  fixed in the limit $\sql\gg 1$.
The highest power of charges in ${1 \ov (\sql)^n}$ term  here is $n+2$.
This  follows, e.g.,  from dimensional analysis, from the fact that higher order terms 
in 2d anomalous dimension operator may contain higher derivative operators (e.g., $E^2$ comes from 
$SO(2,4)$ Casimir   originating from Laplacian  on $AdS_5$, etc.; see \cite{t03}) and also from the fact 
that, in any theory, an $(n+1)$-loop Feynman graph renormalizing a (vertex) operator contains at most 
$(n+2)$ Wick contractions with fields in the (vertex) operator and thus contributes to its dimension 
terms like $Q^{m}/(\sql){}^{n}$ with $m\leq n+2$.

More explicitly, if we consider a string state  with an orbital momentum $J_3 \equiv J$
in $S^5$  and one extra oscillator number $N$ (corresponding, e.g., to an intrinsic 
 spin component due to an extended nature of the string) 
 which determines the value of an  effective string level 
 then \rf{1} is  a consequence of the 
   following  
   2d marginality condition\foot{Here the $(-E^2+J^2 + ...)$ term is 
   the 1-loop  correction to the 2d (anomalous)   dimension, 
 the next term is the  2-loop correction, etc., 
 with  all the terms at the same  order in ${1 \ov \sql}$ being here  on the same
footing. 
This expansion   should    emerge in the sigma model approach upon
  diagonalization of the 2d anomalous  dimension  matrix 
   (as, e.g.,  in the NSR approach
or in the context of a pure spinor approach
 like the one discussed  in  \cite{ps}).
  Here  we ignore  possible shifts 
 of $N$ and $E$  by integers that depend on  a choice of a reference 
 vacuum state (in the bosonic string  context the l.h.s. of \rf{01}
 should be equal to 2).}
\be 
&& 0= N +{1 \ov 2\sql} ( -E^2 + J^2  +n_{02} N^2 + n_{11}   N ) \no
 \\ && \ \ \ \ \ \ 
\ \ + \ {1 \ov 2(\sql)^2 } (  n_{01}   N J^2  +  n_{03} N^3  + n_{12} N^2 +  n_{21}   N  ) 
+ O({\textstyle {1 \ov (\sql)^3 }} ) \ . \la{01}
\ee
Including also some higher-order terms,  the 
 resulting expression for $E^2$  may be  written as\foot{Here 
the coefficient of $J^2$  in the first line should 
  be 1  to be consistent with the BMN limit 
 $N=0$. Again,   we assume that in general 
  $E$ and $J$   may  be  redefined    by possible constant
 shifts    to be consistent with   positions in a supermultiplet 
 (e.g., 
  $E(E-4) = J(J+4) + ...$ is equivalent to  $(E-2)^2 = (J+2 )^2 + ...$   
  for simplest point-like states). This depends on a definition of string vacuum, 
  see \cite{rt2} for more details.} 
  \def \noo {{\tilde n_{01}}}
\be 
&& E^2 = 2 \sql   N  +  J^2 + n_{02}  N^2 +  n_{11} N\no \\ &&
\ \ \ \ \ 
+ \  {1 \ov \sql} \big( n_{01}  J^2 N   + n_{03} N^3 + n_{12} N^2 + n_{21}  N  \big) \no \\ &&
\ \ \ \ \ 
+ \  {1 \ov (\sql)^2} \big( \nm  J^2 N
+   \tn_{02}  J^2 N^2   +n_{04} N^4 +n_{13} N^3 +n_{22} N^2 + n_{31} N\big) \no \\
&&
\ \ \ \ \ +  {1 \ov (\sql)^3} \big(\noo  J^4 N + 
 \tn_{21}  J^2 N +  
 \tn_{12}  J^2 N^2   +  n_{05} N^5+...\big) \no \\
 &&
 \ \ \ \ \ +  {1 \ov (\sql)^4} \big(\ttn  J^4 N +...\big) + 
O({\textstyle{1 \ov (\sql)^5}} ) 
\ . \la{2}
 \ee
 This expression  follows  under the assumption that in \rf{01}  $E^2$ enters only in the 1-loop $1 \ov \sql$ 
 term.  On general grounds, as $E$ may be thought of as a global charge  analogous to $J$, 
 one might wonder if  \rf{01}  should also contain terms like 
 ${1 \ov (\sql)^k} (   E^{k+1} + ...+  E^{m} N^n  + ...) $. 
 However,  terms  depending only on $E$  (or on $E$ and $J$) 
 should   be 2d scheme-dependent 
 (like higher powers of Laplacian in 2d anomalous dimension operator)  and 
 would also  contradict BMN limit $E=J$ in the absence of other charges ($N=0$)
 leading to spurious $1 \ov \sql$ dependent solutions  of the marginality condition;
 they should thus be absent in a scheme preserving target space supersymmetry. 
 Terms  in \rf{01}  involving both   $E$ and $N$ like  ${1 \ov (\sql)^k}  E^n N^{m}$ with $ m+n\leq k+1$, 
 may be present, but in solving the 
 marginality  condition \rf{01} for $E$  in perturbative  expansion in $1 \ov \sql$ 
 they cannot modify the leading-order solution $E^2 = 2 \sql N + ...$ and 
 their 
 perturbative treatment  leads just to redefinitions of coefficients already present in eq.~(\ref{2}). 
 Note also that the presence of the mixed terms $J^k N^m$ terms  reflects the fact 
 that in curved space the center-of-mass  and internal degrees of freedom do not in general decouple.

 Expanding \rf{2} in large $\sql$ for {\it fixed}
  $N,J$ we get 
\be 
&& E= \sqrt{ 2 \sql   N} \Big[ 1 + {A_1 \ov \sql} + { A_2 \ov (\sql)^2} + 
 { A_3 \ov (\sql)^3} + O({\textstyle
 {{1 \ov (\sql)^4} }})\Big] \ , \la{3}\\ 
&& A_1 = {1 \ov 4 N}  J^2 +{1  \ov 4} ( n_{02}  N +  n_{11} ) \ , \ \ \ \ \ \la{4e}\\
&&A_2 = - {1 \ov  2 }  A_1^2  +{1 \ov 4}( n_{01}  J^2    + n_{03} N^2 + 
n_{12} N + n_{21}    ) 
  \no \\
&& \ \ \ \ =  {1 \ov  4 }\Big[n_{21} - {1  \ov 8} n_{11}^2  
+  (n_{12}  -  {1  \ov 4} n_{11} n_{02}    ) N   
  + (n_{03}  - { 1 \ov 8}  n_{02} ^2    )    N^2\Big]  + O(J^2)\ , 
\la{4}\\
&&
A_3= { 1 \ov 128} \Big[   ( n_{11} ^3 - 8 n_{11} n_{21} + 32 n_{31}) + 
 (3n_{02} n_{11}^2   - 8n_{11} n_{12}  - 8n_{02} n_{21}  
 + 32 n_{22} ) N \no \\
 &&\ \ \ \ \ \ \ \  \ \ \ + 
  (3  n_{02}^2n_{11}  - 8 n_{03}n_{11}  - 8n_{02}n_{12}    + 32 n_{13}) N^2 + ...
\Big] \ . \la{4m}\ee 
Substituting particular values of $N$ and $J$ into \rf{2},\rf{3} 
 one can find the 
expansion of the corresponding  quantum string state energy, i.e. the 
strong-coupling
expansion of  the dimension  of the dual 
 gauge-theory operator. 
 Note that the first two terms in the r.h.s. of \rf{2} have direct flat-space 
interpretation, so that $N$ plays the role of string level and the spinning 
string states  with  maximal value of $N$ for a given value on spin belong to the leading
Regge  trajectory. 
 For example,  $N=0$ corresponds to massless (supergravity)
 states  and $N=2$ to states on the first excited string level
 which  contains the Konishi long multiplet as its ``floor''  and  also its ``KK  descendants'' 
 with higher values of $J$  obtained by tensoring with $[0,J,0]$  representation  \ci{ber}.
 The   states in the  Konishi multiplet that we will consider here  correspond to
 $N=2,\ J=2$, see  \cite{rt1,rt2}.


The goal  is thus to determine the coefficients $n_{km}$ in \rf{2}. 
To achieve this one may use the observation \cite{rt1,tt}  that   a 
similar expansion of the string energy  is 
also found 
by  starting  with  a solitonic string  carrying the same 
types of charges as the vertex operator representing a particular quantum string
state  and 

\noindent 
 (i)  first performing 
  the semiclassical expansion 
$\sql \gg 1$  for 
{\it fixed} 
charge densities 
$\Q_i= {1 \ov \sql} Q_i$, 
i.e. $(\N,\J) = {1 \ov \sql} (N,J)$, and then \ 

\noindent
(ii) expanding $E$ 
 in small values of $\Q_i$. Indeed, 
the limit  $\Q_i= {Q_i \ov \sql}  \to 0$  should
 correspond to taking  ${ \sql}\gg 1 $ for fixed
 values of the quantum charges $Q_i$. 
 Assuming that there is no order of limits problem, 
 the same coefficients  $n_{km}$  should be found in these
  two different approaches.
  
 Writing \rf{2}  in terms of $ \N,\J$ as 
 \be 
&&\left({ E \ov  \sql}\right)^2 =  2\N  +  \J^2 +
n_{01}  \J^2 \N  + n_{02}  \N^2  + n_{03} \N^3 + n_{04} \N^4 + \noo  \J^4 \N+ {\tilde n}_{02}\J^2\N^2
+ ...
\no \\ &&
\ \ \ \ \ 
+ \  {1 \ov \sql} ( n_{11} \N +  \nm  \J^2 \N   + \ttn \J^4 \N  + n_{12} \N^2 + \tn_{12} \J^2 \N^2
+ n_{13} \N^3+ ...) \no \\ &&
 \ \ \ \ \ 
+ \  {1 \ov (\sql)^2} ( n_{21}  \N  + \tn_{21} \J^2 \N   +    n_{22}  \N^2 + ...)  +\ost
\ ,   \la{04}
 \ee
one can then  interpret the coefficient  $n_{km}$  in \rf{2} 
as a 
$k$-loop contribution to a term scaling as $\N^m$ in the semiclassical expansion, i.e. 
 $n_{0m}$ can be  extracted from the classical string energy, 
 $n_{1m}$ -- from the 1-loop semiclassical correction, etc.
 Expanding $E$ in \rf{04} in small $\N$ for fixed $\J$  we get
 \be 
 &&{E \ov \sql}= \J+   \Big[  { \N \ov  \J}(1  + \ha n_{01} \J^2 +\ha \noo \J^4+ ...)
 \no \\ 
  && \ \ \ \ \ \ \ \ \ \ \ \ \ - { \N ^2 \ov 2 \J^3}  \Big(1 +   (n_{01} -  n_{02})\J^2 +
    (\noo-{\tilde n}_{02} + {\textstyle{1\ov 4 }} n_{01}^2) \J^4  + ...\Big) + ...\Big]
   \no \\
  && \ \ \ \ \ \ + {1 \ov \sql} \Big[
 { \N  \ov 2 \J }( n_{11}   +  \nm  \J^2+\ttn  \J^4+  ...) \la{05} \\
   &&\ \ \ 
  + { \N ^2 \ov 2 \J^3} \Big(  -  n_{11}  + 
  (n_{12}-\ha n_{01} n_{11}  -     \nm) \J^2
   + (\tn_{12} - \ttn  -\ha n_{01}\nm - \ha  \noo n_{11})  \J^4+...  \Big) 
   \nonumber\\
   &&\ \ \ \ 
    + \frac{\N^3}{4\J^5}\Big(3n_{11}+ [3 {\tilde n}_{11} - 2 n_{12} +
   ( 3 n_{01}  - n_{02} ) n_{11} ]\J^2
    \nonumber\\
   && 
    +\big[  2(n_{13}-{\tilde n}_{12})-n_{01}n_{12} + 3 \ttn 
    + (3{\tilde n}_{01}-{\tilde n}_{02}  + \textstyle{\frac{3}{4}}n_{01}^2    ) n_{11}
    + (3{n}_{01}-{ n}_{02}) \tn_{11} 
 \big] \J^4+...\Big) + ... \Big]
 \no \\  &&
   \ \ \ \ \ \ \ +   {  1 \ov (\sql)^2} \Big[ {\N \ov 2  \J } \big(  n_{21} + \tn_{21}\J^2+ 
   ...\big) + ...
   \Big]    + \ost \ . \no
 \ee
It should be  noted that the quantum string sigma model 
 loop (i.e. $\a' \sim {1 \ov \sql}\ll 1 $)
  expansion in \rf{2}   is  of course different from the semiclassical loop expansion in 
  \rf{04}: in \rf{01} or \rf{2}  the first order  $N$  term is classical, 
  $J^2 + n_{02}  N^2 +  n_{11} N$ are 1-loop  terms, etc., i.e.
  the  coefficients $n_{km}$, 
   in general,  appear at different loop orders in the two expansions.\foot{Note  
   that $n_{\ell 1}$ ($\ell=1, 2, ...$)  are still
    $\ell$-loop coefficients  in both expansions.}
  Note also that while  each $\ell$-loop  term in \rf{2} is a 
  polynomial  of {\it finite} degree, ($\ell+1$),  in the charges, 
  this does not in general apply to the semiclassical expansion \rf{04}
  where each term may contain an  infinite series of terms 
  in the small $\J, \N$ expansion. 
  To relate the two expansions one would need  to reorganize or even resum 
   them.\foot{In particular, 
   considering $ \J \gg \N$   expansion will lead to inverse powers
   of $\J$ in the semiclassical expansion and thus will require a resummation to relate 
   it to \rf{2}.}
   For example, the classical string energy term in  \rf{04} 
  receives contributions from all higher
  loop orders in \rf{2}, etc.\foot{Note also  that ``non-analytic''
   terms \cite{rt1} 
 like $B_2,B_3, ... $  in the large $\sql$ expansion of the energy
 
 \noindent$E= 
 \sqrt{ 2 \sql   N} \big[ 1 + {A_1 \ov \sql} + { A_2 \ov (\sql)^2} +...\big]
 +  B_1 +  {B_2 \ov \sql} + {B_3 \ov (\sql)^2 }+ ...$, 
 which  a priori could   be present in the energy found by using semiclassical expansion, 
 should  not actually appear if this approach is consistent: 
 they would lead to $\fl$ dependent terms in $E^2$, i.e. 
 $E^2 = 2 \sql   N + 2 \sqrt{ 2   N} \big[ { B_2   \ov \fl} +
  { B_3   \ov (\fl)^3}+ ...\big] + ...$
  which  cannot  be present  in the standard sigma model
   perturbative computation of  eigenvalues of 2d 
  anomalous dimension matrix.
  }

 Comparison of \rf{05} or \rf{3}--\rf{4m} to \rf{2} shows that 
 eq.\rf{2}  for the square of the energy    provides a much more ``economical'' description of the
 spectrum. Computing the semiclassical expansion \rf{05} directly one 
 finds indeed many relations between the coefficients there in agreement 
 with the general structure of $E^2$ in \rf{2}.

The expression for $E^2$ in \rf{2} or in \rf{04}
 may  be formally organized as an expansion in  small $\N$  which will  then 
    look  like  an expansion  in powers of  $N$:
  \be \la{5}
E^2 =J^2 +   h_1 ( \l, J) N  +  h_2 ( \l, J) N^2     +  h_3( \l, J) {N^3 } + ... \ , 
\ee
where  for fixed $J$ and large $\l$ the coefficient functions $h_k$ are given by 
\be
&& h_1= 2 \sql + n_{11} + { n_{21} \ov \sql}  + {  n_{31}  \ov (\sql)^2} + ...
 + J^2 \big( {n_{01} \ov \sql}  + {\nm  \ov (\sql)^2}  + {\tn_{21} \ov (\sql)^3}   +...
 \big) + ... 
 \ , 
\la{3k}\\
&& h_2=  n_{02} + { n_{12} \ov \sql}  + { n_{22} \ov (\sql)^2} + ... + 
J^2 \big( {\tn_{02} \ov (\sql)^2} +  {\tn_{12} \ov (\sql)^3}+ ...\big) + ...  \la{3n} \ , \\
&& h_3=  { n_{03} \ov \sql}  + { n_{13} \ov (\sql)^2}+ ...  
\la{3m} , \  \ \ \ \ \ \ \ \ \ 
 h_4=  { n_{03} \ov (\sql)^2} + ...  \ . 
 \ee
 The corresponding expansion of $E$ in small $\N$ for fixed $\J$  is then 
 \be 
 E= J + {  1 \ov 2 J} h_1 ( \l, J) \, N + ...\ ,  \la{555}
 \ee
 i.e. ${  h_1 ( \l, J) }$ may be called, following \cite{bas}, a ``slope'' function.
 In ref. \cite{bas} it was found exactly   in the case of the folded string
 with spin $S$ in $AdS_5$ (in this case  $N=S$). 
 While the coefficients in the ``slope'' function
 $h_1$ are expected, by analogy with the case in  \cite{bas}, 
 to be rational  ($h_1$ is determined \cite{bas}  by the asymptotic Bethe ansatz 
 and is also not sensitive to the phase) the coefficients in the 
 next ``curvature'' 
 function $h_2$  are already transcendental  (as we shall discuss below 
 $n_{12}$ contains $\zeta_3$, $\tn_{12}$ contains $\zeta_5$, etc) 
 and $h_2$ is expected to   be sensitive to ``wrapping'' corrections.

 \subsection{Summary of results for the  coefficients}

 Below we shall consider the examples  of ``small'' semiclassical spinning 
 string states 
 discussed in  \cite{rt1,rt2} that fall into the class of states 
 described   by \rf{2},\rf{04},\rf{05}.
 They correspond to quantum string states  with angular momentum $J$ and few oscillator
 modes excited that are responsible for non-zero components  of intrinsic spin.
  More specifically, we shall consider and compare  the following 
   solutions:\foot{We shall use the following
  notation: $S_1$ and $S_2$  will stand for spins in $AdS_5$; 
  $J_1\equiv J'$ and $J_2$ will be spins in $S^5$ and $J_3\equiv J$ will be orbital momentum in
  $S^5$.}
  two folded string cases: $(S,J)$  and $(J',J)$  and three rigid two-spin circular string
  cases:  $(J_1=J_2\equiv J', J)$, $(S_1=S_2\equiv S, J)$ and $(S=J_1\equiv J', J)$. 
  For lowest values of the winding numbers these represent (in the flat space limit)
  states on the leading  Regge trajectory with the string level  being $N=S$ or $N=J$ in 
  the folded one-spin cases and  $N= 2J'$  or  $N= 2S$  in the circular two-spin cases.
  
 For example, for $N=2$ these represent  states on the  first excited string level. 
 In this case all states with fixed $J$ (i.e. on  a fixed KK level \cite{ber})
 should  belong to a  single long $PSU(2,2|4)$ multiplet.\foot{
 For example, the three circular string  states 
 in the flat space limit are related by Lorentz transformations  and thus belong to the
 same multiplet.
 This  should remain  so upon switching on the curvature.}
  Furthermore, the string states with 
   $N=2,J=2$  are dual to  particular states in the Konishi multiplet on the gauge
 theory side \cite{rt1,rt2}.

As all operators  in a  given supermultiplet should have 
the same 4d anomalous dimension, that means that the corresponding 
string states  should have  the same 
 target space  energy (up to constant integer or half-integer shifts
reflecting their positions in the supermultiplet; such shifts  are ignored in \rf{2}), i.e.
  the expression for $E_{{N=2}}$ as a function of 
  $J$ and $\l$  should be universal, with  $E_{_{N=2}}(J=2, \l)$
  being equal to the dimension of the Konishi multiplet.

 As follows from \rf{2},  this expected  universality of the $N=2$ value 
 of the energy for   any $J$ and $\sql$ imposes  the following 
 invariance constraints 
 on the coefficients of states within a supermultiplet: 
 \be 
&&n_{01}= {\uni} \ , \ \ \ \ \ \ \ \ 2 n_{02}  +  n_{11} = {\uni} \ , \ \ \ \ \ \ \ \ \ \
 4 n_{03}  + 2 n_{12} + n_{21} = {\uni} \ , \la{uni}
 \\ 
 &&   2 \tn_{02} + \nm   =\uni  \ , \ \ \ \ \ \ \ \ \ \
  8 n_{04}  + 4n_{13}  +2 n_{22}  + n_{31} = \uni  \ ,\ \ \  \ \ ... \la{ver}
 \ee
 Note that these conditions  relate different terms in the semiclassical loop expansion. 
 Once the values of these coefficients are known at least for one 
 state in the multiplet, then \rf{uni},\rf{ver} constrain  the coefficients 
 for other states. 
 
 Explicitly, these universal coefficients enter $E_{{N=2}}$ in \rf{2},\rf{3} as follows
 \be 
&& E_{_{N=2}}= 2 \fl \Big[ 1 + {a_1 \ov \sql} + { a_2 \ov (\sql)^2} + 
{ a_3 \ov (\sql)^3}
+
O({\textstyle{
 {1 \ov (\sql)^4}
}})\Big] \ , \la{44}\\ 
&&
a_1= (A_1)_{_{N=2}}= {1 \ov 8} J^2  + {1 \ov 4} (2 n_{02} +  n_{11}) \ , \la{441} \\
&&
a_2 = (A_2)_{_{N=2}}=- {1 \ov 2} a_1^2  + {1 \ov 4} n_{01} J^2 
+ {1 \ov 4} ( 4 n_{03}  + {2} n_{12} + n_{21})  \ , \la{442} \\
&&
a_3 = (A_3)_{_{N=2}}=  - a_1 a_2 +  {1 \ov 4} (2 {\tilde n}_{02} + {\tilde n}_{11})J^2 
 + {1 \ov 4} (  8 n_{04}+ 4 n_{13}+ {2} n_{22}+  n_{31}) 
\ . \la{443}
\ee
$(a_k)_{J=2}$ are then the coefficients of the string coupling expansion of the dimension 
of the Konishi multiplet. 
 $a_1$ thus depends on tree-level $n_{02}$ and 1-loop $n_{11}$
coefficients; $a_2$ depends  on tree-level, extra  1-loop $n_{12}$
and  also 2-loop $n_{21}$ 
coefficients;  $a_3$ depends  on tree-level, extra 1-loop 
${\tn}_{11},n_{13}$, extra 2-loop $n_{22}$  and also 3-loop $n_{31}$
coefficients, etc. 

In general, the highest loop order $\ell$ coefficient 
$n_{\ell 1 }$ in $a_\ell$  originates from the  slope function $h_1$ in \rf{3k}
and thus should be {\it rational} (as was found for the $(S,J)$ folded string state in
 \cite{bas}).\foot{In particular,  
  for the $(S,J)$ folded string state  \cite{bas}:  
$n_{11}= -1, \ n_{21}= -{1 \ov 4}, \ 
n_{31}= -{1 \ov 4},\ n_{41}= - { 25 \ov 64},\ 
 n_{51}= - { 13 \ov 16},\   n_{61}= - { 1073 \ov 512}, $         etc.}
 The subleading loop order coefficient $n_{\ell-1,  2 }$ (for $\ell > 1$) 
 originating from $h_2$ in \rf{3k} should   already be  {\it transcendental}
 -- containing  zeta-function $\zeta(2\ell-1) \equiv   \zeta_{2\ell-1}$. 
 Also,  $n_{\ell-2,  3 }$   (for $\ell >2$) should contain 
   $\zeta_{2\ell-1}$, etc. 
 Then the highest transcendentality term  in $a_\ell$ 
 in \rf{44} should  contain  $\zeta_{2\ell-1}$. 
 
 Indeed, 
 as we shall see  below the 1-loop coefficients $n_{1k}$ 
 obey this pattern: $n_{12}$ contains  $\zeta_3$, $n_{13}$ contains  $\zeta_5$, etc.
 What is unclear at the moment is if the 2-loop and higher coefficients in $h_2, h_3, ...$ (like 
 $n_{22}, n_{32}, ...$)  may contain  other
  transcendental constants as well.\foot{For
 example, the  2-loop and higher order terms in the $\ln S$ coefficient of the
  large $S$   limit of the 
 folded string energy expanded in $1 \ov \sql$ 
  contain  Dirichlet beta function  constants 
 ${\rm K}= \beta(2)$, etc.  (as well as $\zeta_k$) \cite{rt22,bkk}.}
 It would be important to carry out an  explicit 2-loop  computation of $n_{22}$ to 
 clarify this question. 
 
 It is interesting to note that the  weak-coupling  expansion of the anomalous 
  dimension 
 of the Konishi multiplet states also 
  contains $\zeta_k$ constants at  4 and 5  loops  (see, e.g., 
 \cite{eden} and refs. there)  while  the transcendentality origin of higher
  loop coefficients here  again appears to be an open question
  (an answer  should follow from an 
  analytic solution of TBA equations at weak coupling 
  \cite{gkv,f}).

 Let us now  summarize what is  known  \cite{rt1,gssv,rt2,bm,bas,gva}
 and what will be found below 
  about the coefficients $n_{km},\tn_{km}$ in \rf{2} 
  using   the  semiclassical approach. 
   We  will 
  try to identify the   general universality 
  patterns in the structure of   these  coefficients.   
   First, in  all cases 
\be  n_{01}=1  \ , \ \ \ \ \ \ \ \ \   \noo=-{1 \ov 4} \ .  \la{6} \ee
The universality of $n_{01}$ is in agreement with \rf{uni}. 
 This  follows from the universal form of the 
  ``near-BMN''  expansion of the classical string energy: 
\be\la{7}
E^2 = J^2 + 2 N \sqrt{ \l  + {J^2}} + ...= J^2 +  N\Big( 2\sql  
+   { 1\ov \sql} J^2  - {1\ov 4
(\sql)^3} J^4 +  ...\Big) + ... \ , 
\ee
where we assumed that $ N \ll  J \ll \sql$. In other words, the  first term in the semiclassical
expansion of the slope function $h_1$ in \rf{5} is universal:
$ 
h_1 (\l, J) = 2 \sql \sqrt{ 1 + \J^2} + O(\J) $.

The  classical $n_{02},n_{03}$ and the leading 1-loop  $n_{11}$  coefficients are also
rational  \cite{rt1,gssv}. 
We find that in all cases 
\be 
2 n_{02} +  n_{11} = 2   \ , \la{4w}
\ee
verifying the first universality relation in \rf{uni}.
 The value of  $\nm$ is  determined by the term linear in $\N$ in the 
1-loop semiclassical energy computed for fixed $\J$ and small $\N$
and then expanded in small $\J$ (see \rf{05}).
The results  for the folded string \cite{bas,gva} and the 
circular string results described below   imply  that in all cases 
\be \la{333}
\nm= - n_{11} \ , \ \ \ \ \ \ \ \ \ \ \ \ttn = n_{11} \ .  \ee
More generally,  these   results 
imply    the  universality (for the states on the leading Regge trajectory) 
of the $\J$-dependence of the 
first two  leading terms in the  ``slope'' 
function $h_1$ in \rf{5} expanded in the semiclassical limit $\sql \gg 1$ 
 with $\J= { J \ov \sql}$ held fixed:
\be 
 h_1= 2\sql \sqrt{1 + \J^2}  +  \frac{n_{11} }{ {1 +\J^2}}\ +\ {1 \ov \sql} \big[ n_{21} +
 \tn_{21}\J^2+  
 O(\J^4) \big] + 
 O({1 \ov (\sql)^2})   \ .  
 \la{666} \ee
We find also that the 
 leading term in the semiclassical expansion of $h_2$ in \rf{3n}  has 
 the following general form 
\be
h_2=  n_{02} +  { \tn_{02} \J^2 \ov 1 + \J^2 } + {1 \ov \sql}
 \big[ n_{12} + \tn_{12} \J^2  + O(\J^4)   \big] +  \os   \ . \la{75}
\ee
Again, by inspection in all cases we observed, 
in agreement with first relation in \rf{ver} we find, 
\be 2 \tn_{02} + \nm =0 \ ,            \ \la{57}  \ee
so that (using \rf{4w},\rf{333})
\be  
  \tn_{02}= {1 \ov 2} n_{11}=       1- n_{02} \ . \la{58}
\ee
The  1-loop coefficient $n_{12}$ in \rf{2},\rf{75} 
 contains  a transcendental 
$\zet$ part. This   
was first observed  in the small-spin expansion of the folded string
\cite{tt,bd} and pulsating string \cite{bt} energy,
 indicating also that higher-order 1-loop terms
should contain $\zeta_5$, etc., constants. 
The  computation of $n_{12}$ for the circular 2-spin
 string with $J_1=J_2\equiv J'$ ($N=2 J'$) in \cite{rt1} and
for the folded spinning string ($N=S$)  in \cite{gva} led to the exactly same 
coefficient of $\zet$ in $n_{12}$, suggesting its {\it universality}, i.e.
that\foot{The
$\zet$  coefficient is no longer universal for an $m$-folded string \cite{gva} 
but has simple $m^2$ dependence (see also section 2.2 below for
 the corresponding circular
string case).} 
\be n_{12} = n'_{12} - 3\zet  \  , \la{8}\ee
where $n'_{12}$ is a  rational number  depending on a particular string state
on the leading Regge trajectory. 
The  universality  of the $\zet$  coefficient 
  in \rf{8} will be confirmed   below also for the two other examples 
  of the ``small'' circular string solutions: 
  with two equal spins $S_1=S_2$ in $AdS_5$;  with one 
   one spin in $AdS_5$ and one spin $J_1\equiv J'$ in $S^5$ 
   with $S=J'$, \ $N= 2S$  (in ref.\cite{rt1} only $n_{11}$ was computed in these cases).
%

  As was  found in \cite{bas} from the exact computation of the ``slope'' 
   function $h_1$ in \rf{5}
 for the  ``ground-state'' 
 $(S,J)$  state in $sl(2)$ sector   (corresponding to the 
  folded $(S,J)$ string),  the 2-loop coefficient 
 $n_{21}$  is rational and given by\foot{The simplicity  of this  
 coefficient may a priori  be  surprising as it should 
  be given by some 2-loop  world-sheet theory integral
   (with discrete sum over spatial momenta).} 
\be \la{9} n_{21}= - {1 \ov 4} \ .  \ee
In view of \rf{uni}  and the observed 
universality of $\zet$ in  $ n_{12}$ 
 \rf{8}  the rationality  of $n_{21}$  should apply also to other 
 states  under consideration. Indeed, using the 
 values of $n_{03}=-{3 \ov 8}, \ n'_{12}={3 \ov 8}$ \cite{gva}  and \rf{9} \cite{bas}
  for the folded  $(S,J)$ string case the universality of the  third combination in
  \rf{uni} translates into 
  \be 
  4n_{03}  +{ 2} n'_{12} +  n_{21} = -{1}  \ .   \la{4u}
\ee
Remarkably, as we  shall find below, this constraint 
implies the same value \rf{9} for the 2-loop coefficient $n_{21}$
also for  the folded $(J',J)$, circular $(J_1=J_2,J)$ and circular $(S_1=S_2,J)$ 
strings. 
We thus suggest that 
 this value $n_{21}=-{1 \ov 4}$, like the
 value of the $\zet$ coefficient in \rf{8}, 
 should  again be the same  for all the states on 
 the leading Regge trajectory.\foot{The
 universality of this subleading  coefficient in the slope function 
 is supported  by the fact that while $n_{11}$  is sensitive to the curvature of
 subspace where string moves (i.e. it changes sign between the $AdS_5$ and $S^5$ cases) 
 the 2-loop  correction (determining, in particular,  $n_{21}$) depends on
  the square of the
 curvature.}
  This  universality of $n_{21}$  may  help understand how to generalize
the exact result of \cite{bas} for the  function $h_1$ in \rf{5}
to states outside the  $sl(2)$ sector. 
  While  the direct 2-loop computation  of $n_{21}$ 
 is yet to be done for the circular string cases,  the value \rf{9}
 can be indirectly obtained  from the knowledge  of   the 1-loop coefficients 
 by using the expected universality of the  subleading $a_2$ 
 coefficient in the dimension of the Konishi state \rf{442}.
 
 Note that in  view of \rf{8} and \rf{4u}   the coefficients in the Konishi multiplet energy 
 \rf{44} take the following explicit form 
  \be
(a_1)_{_{J=2}}=1\ ,\ \ \ \ \ \   \ \ \ \ \ \ 
 \ \ \ \ \ (a_2)_{_{J=2}}= {1 \ov 4} -{3 \ov 2} \zet  \ . \la{10}
\ee
 The universality of $ (a_1)_{_{J=2}}=1  $, i.e.  the validity of \rf{4w}
not only for the $(S,J)$  folded \cite{gssv} but also  for the  small circular
 string cases  was already verified  in \cite{rt1,rt2}.

Assuming   the universality of the value of $n_{21}$ in \rf{9}
we get from \rf{4u} 
\be 
 n'_{12}  = -{3\ov 8} - 2  n_{03} \ .   \la{4f}
\ee
We shall explicitly confirm this relation (and thus the $n_{21}= - {1 \ov 4}$
prediction) in section 2 
for the circular $J_1=J_2$ and  $S_1=S_2$  cases.
In the case of the  circular $S=J'$ string 
one has $n_{03}=-{1\ov 2}$ and then \rf{4f} implies 
$n'_{12}= {5 \ov 8} $.
The direct computation of $n'_{12}$ in this case will be discussed in 
section 2.4  and Appendix C. As it will be explained 
in section 2.1, the result depends on a choice of a summation 
prescription over the fluctuation frequencies. One particular summation procedure  
discussed in Appendix C leads to  $n'_{12}= {11 \ov 8} $.
 While so far we were   unable to
identify a prescription leading to the value $n'_{12}= {5 \ov 8} $ 
consistent with the universality of \rf{9}, we believe 
it should exist. 
Further support of the universality of $n_{21}$ comes 
from the  folded $(J',J)$  string  discussed in Appendix D where 
we show that in this case    $n_{03}= {1 \ov 8}$ and $  n'_{12}=-{5 \ov 8}$, 
in agreement with \rf{4f}.


The 1-loop result for the $(S,J)$ folded string in 
 \cite{gva}  (in eq. (B.5) there)  and our present 
 results for the circular  and $(J',J)$ folded 
 string cases all lead
  also to the following universal expression 
 for the coefficient $\tn_{12}$  in \rf{2},
 \be
 \tn_{12}= \tn_{12}' + 3 \zet  + {15 \ov 4} \zeta_5   \ , \la{155}
 \ee
 where $\tn_{12}'$ is a rational  number depending on a particular state. 
 Remarkably, like  in the case of $n_{11} = - \nm$ in \rf{333},  
 the  $\zet$ term here is the same as in  $n_{12}$ in \rf{8},  
 up to the  sign. The coefficient $\tn_{12}$ contributes to a higher
 subleading term  $a_4$ in the Konishi dimension \rf{44}. 
 
 The value of  $\tn_{12}$ can be found from the coefficient of the 
 ${1 \ov 2 \sql} \N^2 \J$ term in \rf{05}, i.e. 
 \be 
 \tn_{12} - \ttn   -  {1 \ov 2 } (n_{01}\nm  +    \noo n_{11})
 = \tn_{12} - {{3 \ov 8}} n_{11}    \  \la{552}\ee
 where we used \rf{6}.
 For example, for the $(S,J)$ folded string the result of \cite{gva} gives 
 \rf{155} with $\tn_{12}'= -{27 \ov 16}$. 
 
 The coefficient   $n_{13}$ can be found also by starting with 
     solutions with $J=0$, expanding in small $\N$ and comparing 
     to \rf{3},\rf{4m} (see section 2): 
  $n_{13}$ is  present  in the $N^2$ term in $A_3$ in  \rf{4m}
 which  appears at one  loop 
 order in the semiclassical expansion  (as ${N^2\ov (\sql)^3}= {\N^2\ov \sql}$).
 Our 1-loop results  for the circular strings ($N= 2J'=2S$)
 imply that 
  \be 
  n_{13}= n_{13}'   
 +  n_{13}'' \zet  + {15\ov 4} \zeta_5 \ ,  \la{uu}
   \ee
   where  $n_{13}'$ and $n_{13}''$ are  rational numbers. 
    The coefficient of $\zeta_5$ is  again  
      universal.  
   In the semiclassical expansion of the energy at fixed $\J$ 
   the coefficient $n_{13}$   first  appears 
   in the  ${1 \ov 4 \sql} {\N^3\ov \J}$ term in \rf{05}, i.e. in the combination 
   \be 
   && 2(n_{13}-{\tilde n}_{12})-n_{01} n_{12} + 3 \ttn 
    + (3{\tilde n}_{01}-{\tilde n}_{02}  + {\frac{3}{4}}n_{01}^2    ) n_{11}
    + (3{n}_{01}-{ n}_{02}) \tn_{11} \no \\ 
   && \ \ \ \ \ = (2 n_{13}''-3)  \zet + 2 n_{13}'- 2\tn_{12}' -  n'_{12}
    - n_{11}^2 +  n_{11} \ , \la{tran}
   \ee
   where we first used \rf{6},\rf{8},\rf{9},\rf{58} and then \rf{155}
   and \rf{uu}. Note that $\zeta_5$ terms cancel out in this combination. 
   The absence of $\zeta_5$  in  the coefficient of $\N^3/\J$ term is seen  
       in the expression for the 1-loop energy for  the $AdS_5$ 
       folded string  in  \cite{gva};  
     we will also find that  the same is true   for the folded string in $S^5$ and 
      the three circular string examples.
     As for the  $\zeta_3$  term  in \rf{tran} appearing  in  the
    coefficient  of $\N^3/\J$  in \rf{05},  the  result of \cite{gva} and 
    our results  described  in section 2 
    and Appendix D 
    imply   that it  depends on a particular solution. Thus   
     $n_{13}''$ is  not universal (we shall list its values
      for different solutions below). 
     The results of \cite{gva} in the 
      folded $(S,J)$ string case  lead to $n_{11}=-1$,
     $n'_{12}={3 \ov 8}$, $n''_{13} = {15 \ov 4}$, 
       $\tn_{12}'= -{27 \ov 16}$
     and thus $n_{13}'= -{9 \ov 16}$. 
    
   We expect 
     the 3-loop  slope coefficient $n_{31}$  to be rational  for all states 
     while the 2-loop  coefficient $n_{22}$   to contain 
    only $\zet$ as its highest transcendentality part, i.e. 
    \be \la{2222}
    n_{22}=n'_{22} + n''_{22}\zet \ .\ee
    Then 
      the universality  of the 
    combination $8 n_{04}  + 4n_{13}  +2 n_{22}  + n_{31}$
     in \rf{ver} is  consistent with the universality of 
   the 
    $\zeta_5$ coefficient in  \rf{uu}. Thus the next-order 
    coefficient $a_3$ in the first excited string level state 
    energy  \rf{443}  should contain a
    $\zeta_5$ part.

    Explicitly, as follows from the above  discussion (cf. \rf{57},\rf{uu}) 
      the coefficients 
    in the energy \rf{44}  for the states 
    on the first excited  string  level take the form:
    \be 
  &&  a_1= {1 \ov 8} J^2  + {1 \ov 2}  \ , \la{775}\\
&&  a_2 = - {1\ov 2} a_1^2 +  {1 \ov 4} J^2    -{1 \ov 4}     - {3 \ov 2} \zet    = 
        - {1 \ov 128} J^4  + {3\ov 16} J^2 - {3\ov 8}   - {3 \ov 2} \zet \ , \la{577}\\
&& a_3 =  - a_1 a_2
 +    2n_{04}+  n_{13}   + {1 \ov 2} n_{22}+  {1 \ov 4}  n_{31}\la{431} \\
&& \     =  {1 \ov 4}  a_1 ( 2 a_1^2 - J^2    +  1  )
 +    2n_{04}+  n'_{13}   + {1 \ov 2} n'_{22}+  {1 \ov 4}  n_{31}
 + ( {3\ov 16} J^2 + {3\ov 4} +   n''_{13}  + {1 \ov 2} n''_{22}) \zeta_3 
 + {15\ov 4} \zeta_5
\no
\ee
The universality of $a_3$ implies that  the coefficient 
of $\zet$ and thus  $ n''_{13}  + {1 \ov 2} n''_{22}$ should have state-independent  value. 
 For the folded $(S,J)$ string $a_1,a_2$ in  \rf{775},\rf{577}  appeared  in \ci{gssv,gva}.  
 In  this case the 3-loop coefficient $n_{31}$   can be inferred from the exact expression \rf{4a} 
 for the ``slope'' $h_1$ 
 in \ci{bas}, i.e. $n_{31}= - { 1 \ov 4}$.
 Using also that for folded string solution  $n_{04}= {31 \ov 64}$
 and the value for  $n_{13}$ in \rf{uu}  given by   $n_{13}=-{9\ov 16}  +{15\ov 4} \zeta_3 +{15\ov 4} \zeta_5$ 
    (see \ci{gva} and  \rf{ss})  we conclude that for this state  we should get 
      \be 
 a_3 =  {1 \ov  1024}  ( J^2  + 4) (   J^4  -  24 J^2 + {4 8}  ) 
 +   {11 \ov 32}    +  {1 \ov 2} n'_{22}    +    {1 \ov 2} ({3 \ov 8}   J^2 +  {9 }   +  n''_{22} )  \zeta_3 
 +{15\ov 4} \zeta_5     
\la{463}
\ee
To fix $a_3$  we thus need to know  the 2-loop coefficient $n_{22}$  in $h_2$ in \rf{3n}.
As the folded string is an elliptic solution,  the required direct  2-loop  string  computation 
appears to be   hard. 
It  should be easier  to find $n_{22} $  for the  rational circular  $J_1=J_2$   solution. 
In that case 
$n_{31}$ should  be again rational, while (see \rf{ch}) 
$n'_{13}=-{3\ov 16}, \  n''_{13}=   -{3\ov 4}  $
so that the coefficient of $\zet$ in $a_3$ is $    {1 \ov 2} ({3 \ov 8}   J^2   +  n''_{22} ) $.
The  universality of this  coefficient 
could be checked by an independent computation of $n_{22}$  
by another circular string, e.g., $S_1=S_2$ one. 

It would be  interesting  also  to extend  the numerical TBA analysis in \ci{ff}
to test the universal $J$ dependence of $a_3$  and extract the value of $n_{22}$ 
for the folded string state. The $J=2,3,4$ data in \ci{ff} suggests that $n_{22} \sim -10$.



\

Let us  now    list    the values of few  leading coefficients 
$n_{km},\tn_{km}$ for various folded and  circular  spinning strings
adding  question marks  next to  the  values that were 
not yet derived directly  but are conjectured to be true  on the basis 
of  the universality of \rf{9} (see also table in Appendix E).
For the  folded strings  with one spin $N$ in $AdS_5$ or $S^5$ 
and  an $S^5$  orbital momentum $J$  one finds:

\noindent
$\bullet$ folded string in $AdS_5$ with $(S,J)$,\ $N=S$ \cite{tt,gssv,gva,bas}: 
\be 
&& \ \ \ \ n_{01}=1   \ , \ \ \  n_{02}={3 \ov 2}  \ , \ \ \  n_{03}=-{3\ov 8} \ , \ \ \ \  n_{04}={31\ov 64} \ ,
\ \ \     \tn_{02} = - {1 \ov 2}     \ , \no \\ && \ \ \ 
n_{11}=-1 \ ,  \ \ \  \nm = 1 
 \ , \ \ \  n'_{12}= {3 \ov 8}  \ , \ \ \  n''_{13}= {15 \ov 4}  \ ,
 \ \ \  n_{21}=-{1 \ov 4}  \  ; \la{12}
\ee

\noindent
$\bullet$ folded  string in $S^5$ with $(J',J)$,\  $N=J'$ \cite{bti,bm}:
\be 
&& \ \ \ \ n_{01}=1   \ , \ \ \  n_{02}={1 \ov 2}  \ , \ \ \  n_{03}={1\ov 8}\ , \ \ \ \  n_{04}={1\ov 64}
\ , 
\ \ \     \tn_{02} =  {1 \ov 2} \ , \no \\ && \ \ \ \ 
n_{11}=1 \ ,  \ \ \  \nm =-1  
 \ , \ \ \  n'_{12}=-{5\ov 8} \ ,   \ \ \  n''_{13}= -{3 \ov 4}  \ , \ \ \  n_{21}=- {1 \ov 4} (?) \ . \la{13}
\ee
The value  of $n_{12}$ in \rf{8},\rf{13}  and $\nm$ 
 will be determined below in Appendix D 
following the algebraic curve  approach of \cite{gssv,bm,gva}.

For the  circular strings with two equal  spins in 
$AdS_5$ or $S^5$ and an  $S^5$  momentum $J$
one finds: 

\noindent
$\bullet$ circular string  with $(J_1=J_2, J)$,\ $N=J_1+J_2=2J'$ \cite{rt1,rt2} (see also section 2.2):
\be \la{14}
&&  \ \ \ n_{01}=1   \ , \ \ \  n_{02}=0  \ , \ \ \  n_{03}=0  \ ,   \ \ \  n_{04}=0  \ ,    \ \ \ {\tilde n}_{02}=1 \ ,
 \ \ \      \no \\ && \ \ \ 
n_{11}=2 \ ,   \ \ \  \nm =-2 \ , \ \ \  n'_{12}= -{3 \ov 8}   \ , \ \ \  n''_{13}= -{3 \ov 4}  \ , 
 \ \ \  
n_{21}
= -{1 \ov 4}(?)  \ ; \ \ \ 
\ee

\noindent
$\bullet$ circular string with $(S_1=S_2,J)$,\  $N=S_1+S_2=2S$ \cite{rt1,rt2} (see 
section 2.3  for  $\nm$ and  $n'_{12}$):
\be 
&& \ \ \ \ n_{01}=1   \ , \ \ \  n_{02}=2  \ , \ \ \  n_{03}=-{1}  \ ,   \ \ \  n_{04}=2     \ , \ \ \  {\tilde n}_{02}=-1 \ ,
\ \ \    \no \\ && \ \ \ \ 
n_{11}=-2 \ ,\ \ \  \nm =2  \ , \ \ \  n'_{12}= {13\ov 8} \ ,
\ \ \  n''_{13}= {15 \ov 4}  \ ,  \ \ \  n_{21}
= -{1\ov 4} (?) \ ; \la{15}
\ee 

\noindent
$\bullet$ circular string with $(S=J',J)$,\  $N=S+J'=2S$:\foot{Note that the values 
of 
all coefficients listed here are given by the mean 
average of the values for the $J_1=J_2$ and
$S_1=S_2$  circular strings: symbolically, $ n(SJ)= {1 \ov 2} [ n(JJ) + n(SS)]$.
An intuitive explanation  for this 
may be that since we are considering a near-flat-space expansion
 certain leading coefficients should  be given  just by sums 
 of independent  contributions
of oscillators in different dimensions.
Then to leading order the $AdS_5$ and $S^5$ directions should  contribute  similarly
in the near-flat expansion,   modulo signs due to opposite sign 
of the   curvature.}
\be 
&& \ \ \ \ n_{01}=1   \ , \ \ \  n_{02}={1 }  \ , \ \ \  n_{03}=-{1\ov 2 }  \ ,   \ \ \  n_{04}={3 \ov 4} 
   \ ,  \ \ \  {\tilde n}_{02}=0 \ ,\ \ \      \no \\ && \ \ \ \ 
n_{11}=0 \ , \ \ \  \nm =0 \ , \ \ \  n'_{12}= {5 \ov 8} (?) \ ,
\ \ \  n''_{13}= {3 \ov 2}  \ , \ \ \  n_{21}=-{1\ov 4}(?)  \  . \la{16}
\ee
It is useful also to add the corresponding expressions for the pulsating strings 
with $N$ being the oscillation number (see \cite{bt} and refs. there):\foot{To get 
the required 1-loop coefficients $n_{11}$  it appears that one is to take the fermions in \cite{bt} 
with antiperiodic boundary conditions. The same applies to folded string cases
discussed in \cite{bd,bt}; this removes $\ln 2$ terms  from $n_{11}$ present in 
the periodic-fermion results of \cite{tt,bd,bt}; it remains to see that 
at the end one  establishes the full    agreement with the algebraic-curve
 computation of \cite{gssv}.}
 
\noindent
$\bullet$ pulsating string in $AdS_5$: 
\be \la{17}
&&\ \ \ \   n_{01}=1   \ , \ \ \  n_{02}={5 \ov 2}  \ , \ \ \  n_{03}=- {13 \ov 8} \ ,  \no \\ && 
\ \ \ \  n_{11}=-  \nm =-3 (?) \ , \ \ \  n'_{12}= {23 \ov 8}(?)  \ , \ \ \  n_{21}=-{1\ov 4}(?)  \ ; \ \ \ 
\ee

\noindent
$\bullet$ pulsating string in $S^5$: 
\be \la{18}
&&\ \ \ \  n_{01}=1   \ , \ \ \  n_{02}=-{1 \ov 2}  \ , \ \ \  n_{03}=-{1 \ov 8}  \ ,  \no \\ && 
\ \ \ \ n_{11}= -  \nm =3 (?) \ ,\ \ \  n'_{12}= - {1 \ov 8}(?)  \ , \ \ \  n_{21}=-{1\ov 4} (?) \ , \ \ \ 
\ee
As discussed in \cite{bt}, for $N=2$ the pulsating strings should also represent states 
on the first excited string level, i.e. in particular
 (for $J=2$)  states from the Konishi multiplet.
 With the above values of $n_{km}$  one  indeed reproduces the coefficients in 
\rf{10}.


\

The rest of this paper is organized as follows.
 In the section 2  we first comment on  the general strategy of 
computing  one-loop correction to the energy of classical 
solitons and then use it to evaluate the 
one-loop contributions  to the energy of the three ``small'' circular 
spinning  strings. 
%
The necessary characteristic polynomials are collected, in a 
factorized form, in Appendix~B.
While the solutions with two spins in $AdS_5$ or with two spins in 
$S^5$ yield coefficients $n_{km}$ 
in line with the expectations and patterns outlined above, the rational terms in the result for the  circular string solution with 
one spin in $AdS_5$ and one spin in $S^5$ are found to be ambiguous,
depending on 
 a choice of 
prescription for the summation of the characteristic frequencies. 
In Appendix~C we compute  the
 one-loop correction to the energy 
of the same small circular string solution using the algebraic curve approach 
and find a result  consistent with a particular worldsheet 
summation prescription.
In Appendix A we discuss the structure of the leading terms 
in   the
 slope function $h_1$ \cite{bas}  in the semiclassical expansion. 
The one-loop correction to the energy of 
 folded string with spin  in $S^5$ is found  in Appendix D. 
Appendix E contains table with values of the leading coefficients 
discussed in this paper.

\renewcommand{\theequation}{2.\arabic{equation}}
 \setcounter{equation}{0}
\section{One-loop correction to energy of ``small'' 
circular strings } 

Below  we shall revisit the semiclassical computation of 1-loop
correction to energy of ``small'' semiclassical 
circular strings discussed in \cite{rt1,rt2}  with the aim to extend
 the expansion to next subleading order allowing one to extract the value of the
 coefficient $n_{12}$ in \rf{2},\rf{4} and thus $n'_{12}$ in \rf{8}. 
 In the case of the $J_1=J_2$ string this was already done in \cite{rt1} but we will
 review this case as well for completeness. 

\subsection{General comments on computation of 
one-loop correction }

We will be interested in computing 1-loop corrections to the energy 
of rigid circular spinning strings in $AdS_5 \times S^5$. While these  solutions 
are among the simplest ones being stationary and leading to fluctuation Lagrangian
with constant coefficients   this problem 
(addressed in the past, e.g., in \cite{ft2,ft3,art,ptt,gv2,rt1}) turns out to be 
subtle. Expanding the string action near the solution  and using 
 a static gauge on fluctuations one ends up with a quadratic fluctuation 
 operator $\Delta_2={\rm diag} (K_B, K_F)$ for 8+8 coupled bosonic+fermionic 
   fluctuation modes. Equivalent 
    result for $\Delta_2$  (restricted to ``physical'' subspace)
     is found in the 
   conformal gauge where 2 massless
   bosonic modes decouple and  their contribution is cancelled against the 
   conformal gauge ghost one. Since for all solutions we will consider the 
   target-space time is proportional to the world sheet one, $t= \k \tau$, 
   the 1-loop correction to the target space energy can be found as 
   \be E_1= { 1 \ov \k} E_{2d}  \la{101} \ee   where $E_{2d}$ is 1-loop correction to 
    energy of the world-sheet theory on $R \times S^1$
    
    \noindent
   ($\tau \in ( - {T\ov 2}, {T\ov 2}),  \ T \to \infty, \  \s \in (0, 2 \pi)$). 
    Since in our case $\Delta_2$ has constant coefficients, $E_{2d}$
   can be found  as $ {1 \ov 2 T} \ln \det \Delta_2={1 \ov 2 T} \ln 
   {\det K_F \ov \det K_B} $. 
   Even though $  \ln \det \Delta_2$ is UV finite\foot{See \cite{gv2, LopezArcos:2012gb}
   for discussions of the UV regularization of such determinants.}, 
   the computation of its finite part on 2d cylinder 
   is potentially ambiguous    --  it may  depend  on how 
   individual  fluctuation modes 
   are defined and how their contributions  are combined together. 
    One  complication is that the space of  bosonic 
   fluctuations is multidimensional. 
   Also, the lack of manifest Bose-Fermi 2d symmetry 
   (like world-sheet supersymmetry in the NSR case) implies an extra ambiguity in choice
   of a consistent  regularization. 
   On general grounds, the choice of a prescription for computation of this quantum
   correction  should be governed  by the requirement of preservation of 
   underlying  symmetries of the theory (i.e. conserved charges, including ``hidden''
   ones)  which are ``spontaneously broken'' by a choice of a particular background
   we are expanding around. A practical implementation of this starting directly 
   with the GS    \adss string   action remains a non-trivial task.\foot{Unfortunately, 
   in more 
   complicated 2-spin cases the integrability-based 
   algebraic curve approach  does not appear to help  with the
   problem of ambiguities in the summation over the fluctuation modes.}
   
   To give an example  of possible ambiguities, consider a model where 
   \be 
   E_{2d}= \ha \sum^h_{r=1} c_r \sum_{p_1=-\infty}^\infty \int {dp_0 \ov 2 \pi}
         \ln \Big[(p_0 + a_r)^2 - (p_1+ k_r)^2  + m_r^2 \Big] \la{am} \ . \ee
	Here $p_1$ is an integer momentum in $S^1$ direction 
	and the sum rules $  \sum^h_{r=1}  c_r=0, \  \sum^h_{i=1}  c_r m_r^2 =0$
	ensure that $E_{2d}$ is UV finite. The shifts $a_i$  and (integer) $k_r$
	reflect particular choice of definitions of fluctuation modes. 
If  one  splits the sum over  fluctuations into $h$  separate 2d 
integrals and formally ignores the UV  cutoffs in them
 one may shift the integration/summation
 variables so that  to completely eliminate the  dependence on $a_r,k_r$. 
 However, if one first combines all the contributions into a single  integrand 
 the finite result will depend on $a_r,k_r$.
 
To evaluate similar 1-loop expressions one may choose to diagonalize 
 $\Delta_2$ first  to get  its determinant over ``flavor'' indices 
  as a product over roots of 
 the corresponding characteristic polynomials, 
 $P_{B,F}(p_0)=$``det''$  K_{ B,F} = \prod_{i}  [p_0 - \omega^{(b,f)}_i (p_1)]$.
 One particular prescription for evaluating the resulting integral over $p_0$ is 
 to  first  Wick-rotate it  
 (which is equivalent to $i  \epsilon$ prescription $p_0 \to p_0 -
  i\epsilon$).\foot{It is not clear a priori   why the standard 
  $i \epsilon$ prescription should be preferred  given that 2d Lorentz invariance is
  broken by the background.}
 Then  performing the integral 
  one gets a sum of absolute values  of the characteristic frequencies
 \be 
 E_{2d\ (mod)}= {1 \ov 4}  \sum^{16}_{i=1} \sum_{p_1=-\infty}^\infty
 \Big( |\omega_{i}^{(b)}(p_1)| - |\omega_{ i}^{(f)}(p_1)| \Big) \ .
\label{2.1}
\ee
Alternatively,  one   may
also  treat the worldsheet theory expanded to quadratic order around
the classical solution as a collection of infinitely many coupled harmonic oscillators
(found by expanding  the 2d fluctuation fields in Fourier series in $\s$) 
and  evaluate the corresponding  vacuum energy using the 1-d Hamiltonian 
(operator) quantization method. 
 As was discussed  in \cite{bla,ptt},
 upon a 
diagonalization of the mixing, the contribution of each normal
 mode to the 
energy will enter in the sum  
with a sign $s_i = \pm 1$ determined by a minor of the mixing matrix, i.e. 
in this case we get
\be
E_{2d\ (s)}=  {1 \ov 4}  \sum^{16}_{i=1} \sum_{p_1=-\infty}^\infty
 \Big[  s_{i,p_1}^{(b)}\
   \omega_{i}^{(b)}(p_1)\  - \ s_{i,p_1}^{(b)}\ \omega_{ i}^{(f)}(p_1)  \Big] \ .
\label{BPT}
\ee
 While this expression is equivalent to   \rf{2.1} in some standard 
  simple cases, this need not
 be true in general.\foot{The expression in \rf{BPT}
 may be thought of also as a result of a generalized 
 $i \epsilon$ prescription: $p_0 \to p_0  - i \tilde s_i \epsilon$, with 
 $ s_i \omega_i = \tilde s_i | \omega_i|$, \ $\tilde s_i^2=1$.}
 The computation in 1-d  Hamiltonian quantization  setting may be sensitive 
 to low values of $p_1$ when sign of $\omega_i$  may fluctuate  with $p_1$ 
 and different treatments may correspond to different choices of oscillator vacuum 
 for low (zero) modes.
 At the same time, the signs of sufficiently high 
  mode number terms (i.e. with $|p_1| > n$ = finite
 number)
cannot be sensitive to them. 
 Indeed, since the mixing of modes  is subleading (at most linear) in $p_1$ 
 compared to the free kinetic term,  
  the mixing can be ignored for large $p_1$;  in particular,  
\be
{|p_1| > n }: \ \ \ \ \ \ \ s_{i,p_1}  \omega_{ i} (p_1)   = |\omega_{i} (p_1)| \ .
\label{ab}
\ee
Since the  transcendental ($\zet, \ \zeta_5$, etc.)  terms that may 
appear in the expression for  the 2d energy can originate
solely from   a summation over infinite range of 
 $p_1$  (the sum over any finite
 set of modes can only produce  a rational 
number)  it follows that the {\it transcendental} parts
 of the 2d energy  should  be  controlled by the $ |p_1|\gg 1 $
limit and thus  should {\it not} depend on a sign prescription.  
Moreover,  fluctuations  with   high mode numbers 
have large 2d energy and thus  probe only short 
worldsheet distances.\footnote{Classical scale 
invariance is broken by the background 
so this notion makes sense; ``short distance" 
is measured with respect to the characteristic 
scale of the  background which is set by the parameters of 
the solution.} Their  contribution is  thus less sensitive to 
 details of the classical solution  which is chosen as an expansion point for the 
the worldsheet action  (they will, however, 
be sensitive to the ``topological" features of the solution, 
such as  winding  number). 
We  may then expect that at least some of the coefficients of the 
 transcendental terms in $E_1$ 
 should  be {\it universal}  within a given Regge trajectory
 (parametrised by values of spins with fixed values of windings). 
This explains, in particular, the universality of
 the $\zet$ term in \rf{8} and of the $\zeta_5$ terms in \rf{155} and \rf{uu}.

 The choice of   signs $s_i$   may itself be sensitive 
 to the definition  of the fluctuation modes (related to shifts in fluctuation 
 frequencies or choice of oscillator vacua that may also be different in different gauge
 choices). In general, one expects that the whole summation prescription 
 should be determined by the requirement   that  the target 
space symmetry algebra is correctly realized on quantum string states. 
There are more practical  physical  conditions  that 
are easier to verify, e.g., the vanishing of the 
 one-loop
correction to the energy 
 in the limit in which all charges go to  zero. The one-loop correction  should also  
 vanish  in the limit in 
which the classical solution becomes  supersymmetric
 (in cases where  such limit exists)\foot{Such a requirement
 may seem  inconsistent with the fact that the exact target space energy should contain a
 charge-independent 
term which describes the position of the corresponding state
 in a supersymmetry 
multiplet. However, 
from the perspective of a quantum string state,  this constant term is governed by the 
 fermionic zero 
mode content and should not be accessible semiclassically.}
e.g., one may require consistency with the BMN limit.

Another requirement one may  impose is an analyticity in the smallest
 charge. Indeed, 
in the presence of a large charge one may expect that turning on another charge 
should be smooth; 
that is, the derivative of the energy with respect to the smallest charge 
evaluated at zero  
should not be singular. This translates into the absence in $E_{2d}$ of 
fractional powers of small charges, 
$Q^\alpha$ with $\alpha<1$. 
Such a requirement of the absence of ``non-analytic'' terms  (see \cite{rt1})
turns out to be  consistent with the  structure of the energy \rf{2},\rf{3}
expected to follow  from   the marginality condition for the 
corresponding   vertex operator.

\subsection{Circular string  with spins $J_1=J_2$ and  orbital momentum $J$}

We shall start with the ``small'' circular string in $S^5$ 
described by the following classical solution \cite{ft2,art,rt1,rt2}
($t= \k \tau$, \ $X_k X_k =1$)
  \be
&& X_1+i X_2=\ a\  e^{i (w \tau + m\s)}
~,~~~~
X_3+i X_4=\ a\  e^{i (w \tau  - m\s)}
~,~~~~
X_5+ i X_6= \sqrt{ 1 - 2 a^2}\    e^{i \nu  \tau } \la{h} \no\\ 
&& 
{\cal E}_0^2 = \kappa^2 =\nu^2+  4 m^2a^2 =\nu^2+   \frac{4 m^2\J'}{\sqrt{m^2+ \nu^2}}    \ , 
~~~~~~~~~
w^2 = m^2 + \nu^2\ , \la{hh} \\
&& 
\J'\equiv  \J_1=\J_2= a^2 w
~,  ~~~~~~
\J\equiv  \J_3= 
(1 - 2 a^2)\,\nu \ , \ \ \ \ \ 
\nu=  \frac{\J}{ 1 - { 2\J'\ov  \sqrt{m^2+ \nu^2} } } \ . \no 
\ee
In the limit $a\to 0$ this becomes a short string with small spin $\J'$.  
 $m$ is a winding number which is to be set to 1 to get a state 
 on the leading Regge trajectory. For $\nu=0$ the classical 
 energy has the same expression as in flat space, $ {\cal E}_0=  2 \sqrt{m J'}$. 
Expanding the classical string energy  $E_0 = \sql \E_0$ 
for $ \J' = { J' \ov \sql} \ll 1, \ 
\J = { J \ov \sql} \ll 1$ and assuming $ \J^2 \ll \J'$  we get for $m=1$
  \be
E_0=2\sqrt{\sql J'}\ 
\Big[1+\frac{1}{\sql}\frac{J^2}{8J'}-
\frac{1}{(\sql)^2}\Big(\frac{J^4}{128J'{}^2}-\frac{J^2}{4}\Big)+\dots\Big]  \ . \la{ec} 
\ee 
More generally,   if we expand in small $\J'$ for 
 fixed $\rho^2= \J^2/(4m{\J'})$, we find  
\be
&&E_0 = 2\sqrt{1 + \rho^2} \sqrt{m\sql J'}
\Big[1+\frac{1}{m\sql}\frac{\rho^2 J'}{ 1 + \rho^2} \no \\
&&\ ~~~~~~~~~~~~ +
\frac{1}{(m\sql)^2}\frac{(4  \rho^2 + 
  \rho^4 - 2\rho^6) J'{}^2}{2  (1 + \rho^2)^2}
+...
\Big] \ , \ \ \ \ \ \ \ \rho^2= {J^2\ov 4m\sql {J'}} \ . 
\la{rr1}
\ee
Expanding this further in the limit  $\rho \to 0$ we  get back to \rf{ec} for $m=1$. 
An alternative expansion corresponding to 
 $\J'\ll 1$ with fixed $\J$ (i.e. $\rho \gg 1$) gives (cf. \rf{5},\rf{7},\rf{75}) 
\be
E_0 = J +\frac{2}{J}\sqrt{m^2\l +J^2}\, J'\ - 
\frac{2m^2\l (m^2\l +2J^2)}{J^3(m^2\l + J^2)} J'{}^2+...
 \ .  \la{ess}
\ee
It is useful to perform  the one-loop 
calculation in terms of the two independent semiclassical 
 parameters $a$ and $\nu$.
 We will first expand in small $a$ 
at fixed $\nu$ and then expand in $\nu$. 
An important feature of this 
expansion is that all 1-loop integrals are then regularized in the IR by a 
non-zero value of $\nu$ or $\J$
and therefore $a^2$ and thus the  spin $\J' $  will 
appear  in the 1-loop world-sheet 
 energy    only in integer powers, $E_{2d}= \sum_k f_k a^{2k} $.  
A further  expansion in small $\J$  can then be carried out 
 in the resulting 
coefficients.\foot{Note that for  fixed $\J$  the  small $\J'$ expansions of $a$ 
and $\k$ (over which we are to divide $E_{2d}$ to get $E_1$ in \rf{101}) are given by 
\be
a =  \frac{\J'^{1/2}}{(\J^2 + m^2)^{1/4}}-\frac{\J'^{3/2} \J^2}{(\J^2 + m^2)^{7/4}} +\OO(\J'{}^{5/2})
\ , \ \ \   \
\kappa = \J+\frac{2 \J'\sqrt{\J^2+m^2}}{\J}-\frac{2\J'{}^2m^2(2\J^2+m^2)}{\J^2(\J^2 + m^2)} +\OO(\J'{}^3)
\no \ee
}
Then 
\be
E_1 &=& \frac{1}{\kappa} E_{2d} = \frac{1}{\kappa} \left[f_0(\nu, m)+f_1(\nu, m)\,a^2+f_2(\nu, m)\,a^4+\dots\right]
\no \\
&=& e_0(\J, m)+e_1(\J, m)\,\J'+e_2(\J, m)\,\J'^2+\dots \ .
\label{222}
\ee       
Note that as  the expansion of $\kappa$ or the classical energy \rf{ess} 
 contains
inverse powers of $\J$,  
terms of higher-order in $\J^{-1}$ in $f_i$  contribute to terms of lower order
in the corresponding expansion of  $e_i$. 
Note also that in view of \rf{101} we have 
\be 
E^2 = E_0^2   + 2 \sql E_{2d} +  ...=
  E_0^2   + 2 \sql \Big[ f_0(\nu, m)+f_1(\nu, m)\,a^2+f_2(\nu, m)\,a^4+...\Big]  +
...\ . 
\la{32}
\ee
To compute the 1-loop energy $E_{2d}$ we need the  quadratic fluctuation operators 
$K_{B,F}$ or the corresponding bosonic and fermionic 
characteristic polynomials. They can be 
extracted from \cite{ft3}  and are  listed in  Appendix B.1. As discussed in the previous subsection, 
we need also to choose 
an appropriate definition of $\ln {\det  K_B \ov \det K_F}$
  or a quantization scheme in the Hamiltonian  approach. 
Since in the present case 
 the characteristic polynomials depend on $p_0$ only through $p_0^2$, 
for each mode number $p_1$ 
we have a positive and a negative root which are equal in absolute value. 
In the  Hamiltonian approach it is 
then  natural to define the vacuum energy as a graded sum of the 
positive roots  (cf. \rf{BPT}).  
Such a prescription  gives the same result 
as the  path integral
 approach with  the ``standard''  $i\epsilon$ 
prescription leading to  \rf{2.1}. 
 We then find that 
the one-loop correction to the energy vanishes in the limit $J'\rightarrow 0$. This is 
a required feature  
since for $J'=0$ ($a=0$)  and $J\not=0$ the solution \rf{h} 
reduces to a  BMN geodesic.\footnote{Let us note that 
to  carry out the calculation  in a path integral approach 
in the case  of   $\J=0$ one should write  the $p_0$
integral  as $\int dp_0 \ln { \det{K_B}\ov \det K_F}
 = -\int dp_0 \,p_0 \frac{d}{dp_0}
\ln {\det{K_B}\ov \det K_F} $. This 
integration by parts step here is legal as $\ln {\det{K_B}\ov \det K_F}$ vanishes fast 
enough at infinity. The resulting 
rational function may then be expanded in $\J'$ and integrated without a 
 difficulty.}

Let us   summarize the results for the 1-loop coefficients \rf{222} 
   in the $\J' \ll \J \ll 1 $ expansion. 
Expanding  $E_{2d}$ first in $a$ at fixed $\nu$ and then
 expanding the  result in small $\nu$ we find for the coefficients $f_k$ in \rf{222}
 (for $m=1$):
\be
&&
\!\!\!\!\!\!\!\!\!\!\!\!\!
f_0(\nu, 1)=0\ , 
\quad
f_1(\nu, 1)={2}-\nu^2 +\frac{3}{4}\nu^4+\OO(\nu^6)\ , 
\quad
f_2(\nu, 1)=-\frac{3}{4}-6\zet 
+ \OO(\nu^2)\ .
 \la{2.3}
\ee
Then $e_0(\J, 1)=0$  and 
\be
e_1(\J, 1)=\frac{2}{\J} -2\J+\OO(\J^3)\ ,
\qquad\qquad
e_2(\J, 1)=-\frac{4}{\J^3}+\frac{2}{\J}\big(\frac{5}{8}-3\zet\big)+\OO(\J) \ . 
\label{2.4}
\ee
Comparing this with the general expression for the energy 
 \rf{05} (here $\N = 2 \J'$) 
we conclude    that the resulting 
values of $n_{11}, n_{12},n'_{12}, \nm$ 
are as given in \rf{8},\rf{14}. The values of $n_{11}$ and $ n_{12}$ were already found 
in \cite{rt1}.

We can also find the  exact dependence of  $f_1$ and $e_1$ on $\J$:
\be
f_1(\nu, 1) = \frac{2}{\sqrt{1+\nu^2}}
\ , \ \ \ \ \ \ \ \ \  \ \ \ \ \ \ 
e_1(\J, 1) = \frac{2}{\J ({1+\J^2})} \ . \la{25}
\ee
Then the coefficient of $\J'$ in the energy, i.e.
the   semiclassical expansion  for the corresponding
circular string analog of the ``slope'' \cite{bas} function 
is (see  \rf{5},\rf{555}) 
\be 
 h_1= 2\sql \sqrt{ 1 + \J^2}  +  \frac{n_{11}}{ {1 +\J^2}} + ... \ , \ \ \ \ \ \ \ \ \ \ \ 
 n_{11}= 2 \ . 
 \la{26} \ee
Together with a similar expression found in the $(S,J)$ 
folded string case \cite{bas,gva} 
this provides an evidence of the  universality of the  general expression in \rf{666}.

Note that when  formally expanded in large $\J$, the function 
 $h_1$ in \rf{26}  takes the following form:
$ h_1= 2J   + {\l \ov J} (1 + \frac{2}{ {J}}  + ...)+ ... $.
Here the $\frac{2}{ {J}}$ term  is different by a 
factor of 2 from the  result for the leading 1-loop finite size correction 
found in   \cite{fk1}. 
This disagreement  should not, however,   be   surprising as the two expansions 
are derived in  different limits (see also Appendix A). In the  present 
case, relevant for ``short'' strings, we assumed that $\J' \ll 1$ and $\J$ 
is fixed.  In contrast, the finite size correction 
calculation of \cite{fk1}  assumed the standard ``fast string'' limit of  
$\J'\gg 1$, $\J\gg 1$ with 
$\J'\ov \J$  being   fixed and then 
 taken to be  small.\foot{Let us recall the  distinction between the 
``small'' and ``large'' circular 2-spin solutions \cite{ft2,ft3}.  The 
distinction is sharp at $\J\equiv \J_3=0$: (i)
 the  solution is ``small'' if  $\J_1=\J_2 =\J'$   is such that 
$\J' < \ha $   (here $\J=0$ since $\nu=0$;  this solution is stable); (ii)  
the  solution is ``large'' if  $\J' > \ha$ --
(here $\J=0$ since  $a^2= \ha$; this solution is unstable).  
For 
 nonzero  $\J$   the ``small''  solution    may be defined by requiring that 
 $\J^2 \ll  \J' $;  then its classical energy still starts with 
$\sqrt{ 4 \J'}$     and  thus scales as $\lambda^{1/4}$ for fixed $J'$.
The ``large'' solution is the one with $\J \sim  \J'$  and $ \J \gg 1 $
so that $\E_0 =\J + 2\J'+ { 1 \ov \J} \epsilon ({J' \ov J}) + ...$.  
It is stable if   $\J'  < { 3 \ov 2}  \J$.
While the ``small' and ``large''   cases 
are   smoothly connected for the folded  spinning string,  
that does not apply  to the  circular 2-spin  case as the two expansions 
have different origins ($a\to 0$ and $a\to { 1 \ov \sqrt 2}$).
}

Let us now present the results for the 1-loop coefficients $f_k(\nu,m)$  in \rf{222} 
in the case of  higher winding numbers $ m \geq 1$ (i.e. for states on subleading Regge
trajectories):\foot{The GS fermions here are taken to be periodic for any $m$ 
(see \cite{mik}).}
\be
\begin{array}{c|c|c|c}
       & f_0 & f_1 & f_2 \cr
       \hline
\vphantom{{}^\big|}
m=1& 0    &  {2}-\nu^2+\OO(\nu^4)  & 
-\frac{3}{4}-6\times 1^4\times\zeta_3+\OO(\nu^2)\\[4pt]
\hline
\vphantom{{}^\big|}
m=2& 0    &  {20}-\frac{17}{2} \nu^2+\OO(\nu^4)  &   
-\frac{89}{6}-6\times 2^4\times\zeta_3+\OO(\nu^2)\\[4pt]
\hline
\vphantom{{}^\big|}
m=3& 0    &  {60}-\frac{247}{12} \nu^2+\OO(\nu^4)  &  
-\frac{3357}{40}-6\times 3^4\times\zeta_3 +\OO(\nu^2) \\[4pt]
\hline
\vphantom{{}^\big|}
m=4& 0    &  \frac{376}{3}-\frac{4043}{108}+\OO(\nu^4)  \nu^2 &  
-\frac{263939}{945}-6\times 4^4\times\zeta_3+\OO(\nu^2) 
\end{array}
\label{J1eqJ2_frequency_sum}
\ee
Simple inspection shows that the coefficient of $\zeta_3$ in $f_2$ 
grows like $m^4$. This dependence is
changed, however, after  we express the parameters of the solution in 
terms of the spins, using, in particular,   the relation $a^2=m^{-1} \J'
+{\cal O}(\J^2)$.
The coefficients $e_k (\J, m)$ in \rf{222} are then found to be:
\be
\begin{array}{c|c|c|c}
       & e_0 & e_1 & e_2 \cr
       \hline
\vphantom{{}^\big|}
m=1& 0    &  \frac{2}{\J} -2\J +\OO(\J^3) & 
-\frac{4}{\J^3}+\frac{2}{\J}\left(\frac{5}{8}-3\times 1^2\times\zeta_3\right)
+\OO(\J)\\[4pt]
\hline
\vphantom{{}^\big|}
m=2& 0    &  \frac{10}{\J} -\frac{11}{2}\J +\OO(\J^3) &   
-\frac{40}{\J^3}+\frac{2}{\J}\left(\frac{319}{48}-3\times 2^2\times\zeta_3\right)
+\OO(\J)\\[4pt]
\hline
\vphantom{{}^\big|}
m=3& 0    &  \frac{20}{\J}-\frac{287}{36}\J +\OO(\J^3) &  
-\frac{120}{\J^3}+\frac{2}{\J}\left(\frac{3821}{240}-3\times 3^2\times\zeta_3\right) 
+\OO(\J) \\[4pt]
\hline
\vphantom{{}^\big|}
m=4& 0    &  \frac{94}{3\J}-\frac{2233}{216}\J +\OO(\J^3) &  
-\frac{752}{3\J^3}+\frac{2}{\J}\left(\frac{289367}{10080}-3\times 4^2\times\zeta_3\right)
 +\OO(\J) 
\end{array}
\la{pat}\ee
As in the folded string case \cite{gva}, the coefficient of $\zeta_3$ 
in $e_2$  grows like $m^2$,  supporting the above 
argument for the universality of the  transcendental terms.\foot{ An interesting open question 
 is how the quantum string states corresponding to folded and circular 
  spinning strings with $m >1$  fit into supermultiplets at higher 
  excited string levels.
  Note, however, that 
the  pattern of the $1\ov \J^3$ terms in $e_2$ in \rf{pat}
 appears to be  different from the one in \cite{gva}.}

It is possible to  find  higher orders in the small $\N=2\J'$ expansion 
of the one-loop correction \rf{222} to the 
energy:
\be
\label{hightrJ1J2}
&&E_1 =
\big(\frac{1}{\J} - \J + \J^3+\dots\big) \N \\
&& \!\!\!\!\!\!
+\Big[-\frac{1}{\J^3} + (\frac{5}{16} - \frac{3}{2} \zeta_3)\frac{1}{\J} 
           - (\frac{69}{32} - \frac{3}{2} \zeta_3 - \frac{15}{8} \zeta_5)\J 
           - (\frac{655}{128} + \frac{25}{16} \zeta_3 + \frac{15}{8} \zeta_5
	    + \frac{35}{16} \zeta_7)\J^3+\dots\Big]\N^2 
           \cr
&& \!\!\!\! \!\!
+\Big[
\frac{3}{2\J^5} + (\frac{3}{16} + \frac{3}{2} \zeta_3)\frac{1}{\J^3} 
           + (\frac{41}{32} - \frac{9}{8} \zeta_3)\frac{1}{\J} - 
    (\frac{175}{32} - \frac{33}{8} \zeta_3 - \frac{25}{8} \zeta_5 + \frac{35}{16}\zeta_7)\J+\dots\Big]\N^3+\dots \ .
    \nonumber
\ee
We notice that through $\OO(\N^2)$ order all the transcendental terms are
 the same as in the case of the folded string 
in $AdS_5$ \cite{gva}; we will find them also to be the same 
 for other  two circular string
solutions and 
the folded string in $S^5$.
Comparing to the  general expansion in \rf{05} where the
corresponding coefficient is in \rf{tran} 
 we find then 
the values   of $\tn_{12}$, $n_{13}$  quoted in \rf{155},\rf{uu}
 with   $\tilde n_{12}' = -\frac{57}{16}$, $n_{13}''=  -{3\ov 4}$ and $n_{13}'=-\frac{3}{16}$.

Let us  now present the result for the 1-loop correction to the 
energy in the limit of small $\J'$ and fixed $\rho^2 = { \J^2 \ov 4 m \sql \J'}$.
At fixed $\rho$  and $\J' \ll 1$ 
the relation between the parameters of the solution and the charges is:
\be
\nu&=&2\rho \sqrt{m\J' }
   \Big[1+\frac{2 \J' }{m}
   -\frac{4 \J'{}^2\left(\rho^2-1\right)}{m^2}
+\OO(\J'{}^3)\Big]\ , \no 
\\
\kappa&=&2 \sqrt{m\J'} \sqrt{1 + \rho^2}
\Big[
1
+\frac{\J'{} \rho^2}{m\left(1+\rho^2\right)}
+\frac{\J'{}^2 \rho^2 \left(4 +\rho^2-2 \rho^4\right)}{2 m^2\left(1+\rho^2\right)^2}
+\OO(\J'{}^3)
\Big]\ , \no 
\\
a^2&=&\frac{\J'{}^2}{m}
\Big[
1
-\frac{2 \J'{} \rho^2}{m}
+\frac{\J'{}^2 \left(6 \rho^4-8  \rho^2\right)}{m^2}   
+\OO(\J'{}^3)
\Big]\ . \la{tak} 
\ee
We may  use these expressions and   $f_k$ in 
(\ref{222})  given in 
(\ref{J1eqJ2_frequency_sum}) to find  the  fixed-$\rho$ 
expansion of $E_1$.
Indeed, since $a^2\propto \J'{}^2$ 
contains only positive powers of $\J'$ while 
 $\kappa$ and $\nu$  do not contain inverse 
powers of $\J'$,  higher orders in the small 
$a$ and small $\nu$ expansion cannot affect lower orders.
For $m=1$ we then find 
\be
\la{tuk}
&&E_1=\frac{\sqrt{\J'}}{\sqrt{1+\rho^2}}\Big[
1 + 
\Big(-\frac{3+43\rho^2 +32\rho^4}{8(1+\rho^2)} - {3\zeta_3}\Big)\, \J' 
%
+\OO(\J'^2)\Big] \ . 
\ee
Taking  the limit $\rho\rightarrow 0$  we 
may  read off the value of the coefficient $n_{12}$ in \rf{3},\rf{4}
(here $n_{02}=0$)
\be n_{12} = -{ 3 \ov 8} - 3 \zet \la{ji}\ee 
 which is  in agreement with 
\rf{8},\rf{14}.

It is  possible also to determine the transcendental part of the next terms in 
the small $J'$ expansion of the one-loop energy directly at $J=0$,   
extending the $\rho=0$ limit  of the expression in \rf{tuk} and
showing that this limit can be safely taken in that equation: 
\be
\!\!\!\!\!\!
(E_1)_{\J_1=\J_2=\J',\ \J=0} \ =
  \sqrt{\J'}\, \Big[1+ \big(-\frac{3}{8}-3\zet\big)\, \J'
  + 2\big(-\frac{3}{16} - {3 \ov 4} \zet  + {15\ov 4} \zeta_5 \big)\, \J'^2
  +\OO(\J'^3)\Big] \,. \la{ch}
\ee
Comparing  to \rf{4m} (where the transcendental part 
of the $N^2$ term  is  contained in $n_{13} - {1 \ov 4} n_{02} n_{12}$) 
 we  find the  value of  $n_{13} $ to  be in agreement with  \rf{uu}
 again with $n_{13}'=-\frac{3}{16}$ and $n_{13}''= -{3 \ov 4} $  (here $n_{02}=0$).  
 

\subsection{Circular string  with spins $S_1=S_2$ and  orbital momentum $J$}

Let us now consider the small string with 2 equal spins in $AdS_5$ 
orbiting big circle in $S^5$ \cite{ft2,art,rt1,rt2}
($Y_0^2 +Y_5^2 - Y_m Y_m  =1$):
\be
&&Y_0 + iY_5 = \sqrt{1+2r^2}  \  e^{i \k \tau} \ , \ ~~~  \  
Y_1 + iY_2 = r \  e^{i ( w \tau +  m\s) } \ , \ ~~~ \
Y_3 + iY_4 = r \  e^{i ( w \tau -  m\s) } \ , \no \\ 
&&  
X_1+i X_2=  e^{i \nu \tau}   \  ,\ \ \ \ \ \  w^2=\kappa^2+m^2  \ , \ \ \ \ \ \ \ 
\kappa^2 = 4 m^2 r^2+\nu^2
\ ,  \la{j}\\
&& \E_0=(1+2r^2)\kappa = \kappa+\frac{2\kappa\S}{\sqrt{m^2+\kappa^2}}~, ~~~~~~~~~~
\S= \S_1=\S_2=r^2w~, \ \ ~~~~\J=\nu  \ . \no
\ee
Short string limit corresponds to $r \to 0$  when the solution
 approaches its flat-space limit (for
$\nu=0$). 
The parameter $\kappa$  determined from the conformal gauge condition 
may be written as 
  \be
  \kappa^2=\frac{4m^2}{\sqrt{m^2+\kappa^2}}\S+\J^2 \ .
  \ee
Below we shall consider the case of  $m=1$.
For small $\S$  and   small $\J$ we get 
 the following ``short'' string expansion of the 
 classical energy  ($E_0 = \sql \E_0$):
\be
\E_0 = 2\sqrt{\S}\ \Big( 1 +\S +\frac{\J^2}{8\S }+ ... \Big) \ . \la{eca}
\ee
In the limit of small $\S$ with fixed $\J$  we get 
\be
\E_0 = \J+\frac{2}{\J}\sqrt{1+\J^2}\,\S-\frac{2\S^2}{\J^3(1+\J^2)}+\OO(\S^3)\ . \la{231}
\ee
At small $\S$ with fixed $\rho^2 = {\J^2 \ov 4 \S}$  we find instead 
\be
\E_0 = \sqrt{\S}\Big[
\Big(-\frac{1}{4\rho^3}+\frac{1}{\rho}+2\rho\Big)
-
\Big(\frac{1}{2\rho^3}-\frac{1}{\rho}-2\rho\Big)\,\S 
+
\Big(\frac{1}{\rho^3}-\frac{5}{\rho}-4\rho-2\rho^3\Big)\, \S^2
+\OO(\S^3)
\Big]\la{ko}
\ee
As in the previous 
 $J_1=J_2$ case it is convenient to carry out the 1-loop 
calculation in terms of $\nu$ and $r$ and 
then evaluate the result in the two  limits: (i)  small $\S$ with  fixed $\J$ 
or (ii)  small $\S$ with fixed $\rho$. As in \rf{222}
 the 1-loop correction to the energy may be
written as 
\be
E_1=\frac{1}{\kappa} E_{2d} &=& \frac{1}{\kappa}\left[f_0(\nu, m)+
f_1(\nu, m)\,r^2+f_2(\nu, m)\,r^4+\dots\right]
\no \\
&=& e_0(\J, m)+e_1(\J, m)\,\S +e_2(\J, m)\,\S ^2+\dots \ .
\label{3333}
\ee       
Using the expressions for the characteristic polynomials in Appendix B.2\foot{They 
can be obtained 
from  those in the $J_1=J_2$ case as the two  solutions are related by 
an  analytic continuation effectively  interchanging the $AdS_5$ and $S^5$ parts, 
$a^2 \to - r^2$, $\k \to \nu$, etc.}
  and the ``standard''  choice of summation prescription \rf{2.1} in which we 
  keep unspecified the signs of the terms that vanish in the $r^2 \sim \S\rightarrow 0$ limit
we found that expanding first in $r$ and then in 
  $\nu$  the expansion of the world-sheet energy  $E_{2d}$ 
in \rf{3333} 
contains the following  terms
\be
&& E_{2d}=E_{2d \ \rm low}+E_{2d \ \rm high}\ ,  \no \\ 
&& E_{2d \ \rm low}=
\Big[-\frac{q}{\nu}-\frac{7}{3}+\frac{235}{216}\nu^2+\OO(\nu^4)\Big] r^2
+ 
\Big[\frac{q}{\nu^3}-\frac{1565}{432}+\OO(\nu^2) \Big]r^4+\OO(r^6)    \la{yyy}
\\
&&E_{2d \ \rm high}=
\Big[\frac{1}{3}-\frac{19}{216}\nu^2+\OO(\nu^4)\Big]r^2
+ 
\Big[\frac{2969}{432}-6\zeta_3+\OO(\nu^2)\Big]r^4+\OO(r^6) \ . \la{yy}
\ee
We split the result into the contribution of few  ``low'' modes ($p_1=0, \pm 1, \pm 2$) 
 and the rest of  ``higher''   modes. The coefficient 
  $q$ of the singular in $\nu\to0$  contributions  depends on the 
  signs  $s_{p_1}$ of low fermionic frequencies 
  which vanish at $r=0$ for $p_1=\pm 1$, i.e. 
  $q= 2 +  s_{1} + s_{-1}$.
  There is thus a choice of a sign prescription that ensures the absence of 
  unwelcome singular  terms in $\nu$. 
  %
  The natural value for this coefficient is $q=0$ as the complete two-dimensional energy 
  of the solution, whose 1-loop part is $E_{2d}$ above, is the right-hand side of eq.~(\ref{01}) 
  and is therefore expected to contain only even powers of $\J=\nu$. 
Setting thus $q=0$, the  resulting values of the  coefficients $f_k$ in \rf{3333} are 
\be
f_0(\nu, 1)=0\ , 
\qquad
f_1(\nu, 1)=-2+\nu^2 + O(\nu^4)\ , 
\qquad
f_2(\nu, 1)=\frac{13}{4}-6\zeta_3 + O(\nu^2)\ . \la{fff}
\ee
Using that $\nu=\J$ and 
\be
r^2 = \frac{\S}{\sqrt{1+\J^2}} -\frac{2\S^2}{(1+\J^2)^2}+...\ , 
\ \ \ \kappa =\J+\frac{2}{\J\sqrt{1+\J^2}}\S -\frac{2(1+3\J^2)}{\J^3(1+\J^2)^2}\S^2+...
\ , \label{params_S1eqS2_3}
\ee
it follows that $e_k$ in \rf{3333} are given by 
\be
e_0=0\ , \ \ 
\qquad
e_1=-\frac{2}{\J}+2\J+\OO(\J^3)\ , \ \ 
\qquad
e_2=\frac{4}{\J^3}+\frac{2}{\J}\big(\frac{5}{8}-3\zeta_3\big)+\OO(\J) \ . \la{ee} 
\ee
Comparing to \rf{05} (here $\N= 2 \S$)  we find, in agreement with \rf{8},\rf{15},  that 
in the present case $n_{01}=1$, $n_{02}=2$, $n_{11}=-2$,   $\nm=2$,  and 
\be
n_{12}=\frac{13}{8}-3\zeta_3 \ .
\label{n12_S1eqS2}
\ee
The value of $n_{11}$  was previously found in \cite{rt1}. 
The  value  $n_{12}'=\frac{13}{8}$ is the expected one, i.e. is 
in agreement with \rf{4f},  implying the universality of the  
value of the energy for the corresponding 
(Konishi-multiplet) state with  $J=S=2$ on the 
lowest massive string level.  

As in \rf{hightrJ1J2} we may 
 determine the transcendental part
  of the higher order terms in the small $\S$ expansion
of the energy ($\N=2\S$)\foot{It is interesting to mention that, 
in a small $\nu$ expansion of the coefficient  $ f_2(\nu, 1)$  in $E_{2d}$, 
at $\OO(\nu^0)$ there is only $\zeta_3 $ term 
and at $\OO(\nu^2)$  there is only $\zeta_5$  for both $J_1=J_2$ and 
$S_1=S_2$ cases.  This implies that $ \zeta_3$  in $\tilde n_{12}$
 has the same origin as $\zeta_3$ in $n_{12}$: the only 
difference in its coefficient comes from the expansion of 
$a^4\ov \kappa$ vs $r^4\ov \kappa$.
}
:
\be
\label{hightrJ1J2x}
&&\!\!\!\!\!\!\!\!
E_1=\big(-\frac{1}{\J}+\J-\J^3+\dots\big)\N\\
&&\!\!\!\!\!\!\!\!
+\Big[\frac{1}{\J^3} + (\frac{5}{16} -\frac{3}{2}\zeta_3)\frac{1}{\J} 
+(-\frac{93}{32}+\frac{3}{2}\zeta_3+\frac{15}{8}\zeta_5)\J+\dots\Big]\N^2 + 
\cr
&&\!\!\!\!\!\!\!\!
+ \Big[-\frac{3}{2\J^5} 
    + (\frac{3}{2}\zeta_3-\frac{3}{16})\frac{1}{\J^3} 
    + ( \frac{9}{8}\zeta_3-\frac{41}{32})\frac{1}{\J} 
           + (\frac{363}{32} - \frac{43}{8} \zeta_3 - 5 \zeta_5 - \frac{35}{16}\zeta_7)\J+\dots
	    \Big]\N^3 +\dots \ .
\nonumber
\ee
%
%
Comparing to \rf{05},\rf{tran}  
the  $\OO(\J\N^2)$ term here gives  the value  
of $\tn_{12}$  in \rf{155} with ${\tilde n}'_{12}=-\frac{105}{16}$. 
Together with the absence of $\zeta_5$ at $\OO(\N^3/\J)$, this 
determines $n_{13}$ as quoted in eq.~(\ref{uu})
with 
$n_{13}'=-\frac{85}{16}$ and $n_{13}''=  {15 \ov 4}$.

Next, let us mention    the case of small $\S$ expansion  
for fixed $\rho^2 = {\J^2 \ov 4 \S}$. 
Since the expressions  in  (\ref{params_S1eqS2_3}) contain the exact $\J$ dependence, 
we may get  the corresponding  $E_1$ from $E_{2d}$ in \rf{3333},\rf{fff}
(cf. \rf{tuk})
\be
E_1 = \frac{\sqrt{\S}}{\sqrt{1+\rho^2}}\Big[-1
+\big(\frac{21}{8} +4\rho^2 -3\zeta_3 \big)\, \S  +\OO(\S^2)\Big]\  . \la{tik}
\ee
Taking  the limit $\rho\rightarrow 0$  we 
may  read off again 
the value of the coefficient $n_{12}$ in \rf{3},\rf{4},\rf{n12_S1eqS2}.\foot{Note that in
\rf{tik}
 we have the following combination:
  $n'_{12} -{1\ov 4}  n_{11} n_{02}=\frac{13}{8} + 1 = \frac{21}{8} $.}

Summing  up  the small $\nu$ expansion of the function $f_1(\nu, 1)$
in \rf{fff}  we may    find  the exact form of $e_1(\J, 1)$ in \rf{ee}:
\be
f_1(\nu, 1) =-\frac{2}{\sqrt{1+\nu^2}}  \  , \ \ \ \ \ \ \ \ \ \ 
e_1(\J, 1) = -\frac{2}{\J(1+\J^2)} \ . \la{fe}
\ee
These expressions are  just
  negative   of the corresponding functions in the $J_1=J_2$ case in \rf{25},
  in agreement with the general expression \rf{666} and the opposite signs of the $n_{11}$
  coefficients in \rf{14} and \rf{15}.

One may also perform the computation of $E_1$   by setting $\J=0$ 
directly from the start.\foot{For $\J=\nu=0$ one has  
$
{\kappa} = 2r =2 \sqrt{\S}-2\S^{3/2}+ {9}\S^{5/2}+\OO(\S^{7/2})
$, etc.}
%
While similarly to the $J_1=J_2$ string  case 
the characteristic  polynomials here  depend only on $p_0^2$ and thus
for each mode number there are two roots equal in absolute 
value and opposite in sign, a sign prescription
similar to that of the $J_1=J_2$ case 
in which the one-loop energy is given by the graded 
sum of the positive roots of the characteristic polynomial \rf{2.1}
leads to an unwanted feature: a non-zero  value for $E_{2d}$ in the 
$\S\rightarrow 0$ limit 
(see also  eq.~(3.35) in \cite{rt1}). 
As discussed in Appendix~A of \cite{rt1}, this constant term may be removed 
by a specific reorganization 
of modes together with a change of integration variables, leading  to a cancellation of 
the problematic  term at the 
level of the $p_0$ integrand (so that   a specific 
$i\epsilon$ prescription  was not necessary). 
The same result may be obtained by adjusting the sign of just  one root of each of the two 
 fermionic characteristic polynomials $F_1$ and $F_2$
which for $p_1=\pm 1$ scale as  $\sqrt{\S}$ in the limit $\S\rightarrow 0$:
  their signs 
should be such that 
their contribution adds up to zero.\foot{Interestingly, the only effect of 
this choice is to remove the problematic   term 
and thus to restore the expected  $\S\rightarrow 0$ limit
(all related  higher  integer powers of $\S$ are simultaneously removed).
}
Then the ``low''  modes with 
$p_1=0, \pm1, \pm 2$ contribute to 
the sum over the roots of the characteristic polynomial 
 as:
 \be 
 E_1 = \frac{1}{\kappa}\Big( E_{2d \ \rm low}+ E_{2d \ \rm high}\Big) \ , \ \ \ \ \ \ \ \ 
E_{2d \ \rm low} = -\frac{7 r^2}{3} - \frac{1565 r^4}{432}+\OO(r^6)  \ , \  \la{er}\\
E_{2d \ \rm high}= 
 \sum_{p_1=3}^\infty \Big[  -\frac{4}{p_1(1 - p_1^2)}r^2+
\frac{4 - 17 p_1^2 + 137 p_1^4 - 40 p_1^6}{p_1^3 (4 - p_1^2) (1 - p_1^2)^3}r^4+\OO(r^6)\Big]
 \ . \la{ki}
\ee
Using that $
E_{2d \ \rm high}  =\frac{2}{3}r^2 +(\frac{2969}{216} - 12 \zeta_3)r^4+\OO(r^6)
$
we find 
\be
E_1 = \sqrt{\S}\, \Big[-1+ \big(\frac{21}{8}-3\zeta_3\big)\, \S+\OO(\S^2)\Big] \ , \la{jp}
\ee
which is the same as the $\rho=0$ limit of \rf{tik}.

It is possible also to find the analog of \rf{ch}, i.e. to 
determine the transcendental part of the next terms in 
the expansion of the one-loop energy  of 
the $S_1=S_2$    string at $J=0$, extending  \rf{jp} to next order:
\be 
(E_1)_{\S_1=\S_2=\S,\ \J=0} \ =
  \sqrt{\S}\, \Big[-1+ \big(\frac{21}{8}-3\zet\big)\, \S
  + 2\big(-\frac{59}{8} + {21 \ov 4} \zet  + {15\ov 4} \zeta_5 \big)\, \S^2   +\OO(\S^3)\Big] \ . 
   \la{011}  
  \ee
  Comparing to \rf{4m} 
  we  conclude that the highest 
  transcendental coefficient $\zeta_5$ at the next order 
  is again universal, leading to the expression 
  for  $n_{13} $ in \rf{uu} again with $n_{13}'=-\frac{85}{16}$ and $n''_{13}= {15 \ov 4}$.

\subsection{Circular string with spins $S=J'$  and orbital momentum  $J$ \label{SeqJp}}

The ``mixed'' \adss  circular 
 solution  is described by (we set the two windings equal to 1)
   \be 
&&Y_0 + iY_5 = \sqrt{1+r^2}  \  e^{i \k \tau} \ , \ \ \  \  \ \ 
Y_1 + iY_2 = r \  e^{i ( w \tau +  \s) } \ , \ \ \  \ \  \  \ \   w^2 = \k^2 +1 \ , \no \\ 
&& X_1 + i X_2  = a \  e^{i ( w'\tau -  \s) } \   , \ \ \ \ \ \ 
X_3+i X_4 = \sqrt{1 -  a^2}\ e^{i \nu \tau}   \  , \ \ \ \  \ w'^2 = \nu^2 +1 \  ,  \la{mm}  \\ 
&&
\kappa^2  - \nu^2 = 2r^2+2a^2 \ , 
~~~~~~~~~~~~~~~~~~
r^2w=a^2w'\ , \no \\
&&
\E_0 =\kappa (1 + r^2) \ , \ \ \ \ \ 
~~~\S=r^2w=a^2w'=\J'~,~~~~ \ \ \  \J =(1-a^2)\nu \ . \la{p}
\ee
Note that this solution is ``self-dual'' under the analytic continuation interchanging $AdS_5$ and $S^5$
parts: $ \k \leftrightarrow \nu, \ r\leftrightarrow i a, \ w \leftrightarrow - w'$. 
The parameters $\kappa$  and  $\nu$  may be expressed in terms of the spins 
 by solving the equations
\be
\kappa^2-\nu^2 = \frac{2\S}{\sqrt{1+\kappa^2}} +\frac{2\S}{\sqrt{1+\nu^2}}  
~, \ \ \ \ ~~~~~~~
\J_2=\nu-\frac{\nu \S}{\sqrt{1+\nu^2}} \ . 
\ee
The classical energy has the following expansions 
\be
&&
(\E_0)_{ \S \ll 1, \ \J={\rm fixed}} 
=\J+\frac{2}{\J}\sqrt{1+\J^2}\S-\frac{2}{\J^3}\S^2+\OO(\S^3) \ , \la{ky}\\
&& 
(\E_0)_{ \S \ll 1, \ \rho^2= { \J^2 \ov 4 \S}={\rm fixed}}  = 2\sqrt{1+\rho^2}\sqrt{\S}\Big[1+
\frac{1+2\rho^2}{2(1+\rho^2)}\,\S\no  \\
&& \ \ \ \ \ \ \ \ \ \ \ \ \ \ \ \ \ \  \ \ \ \ \ \ \ \ \ \ \ \ \ \ 
-\frac{5 + 8 \rho^2 + 12 \rho^4 + 8 \rho^6}{8(1+\rho^2)^2}\, \S^2 +\OO(\S^3)\Big]  \ , \la{rre}\\
&&  (\E_0)_{ \J^2 \ll \S \ll 1} 
= 2\sqrt{\S}\ \Big(1 + {1 \ov 2}  \S +\frac{\J^2}{8\S}+... \Big) \ .\la{up}  
\ee

As  in the previous cases we shall carry out the 1-loop computation 
in terms of the parameters $\nu$ and $r$ and then evaluate the result in the small $\S$
limit with  fixed $\J$ or  fixed $\rho^2={ \J^2 \ov 4 \S}  $, i.e. we will define $f_k$ 
and $e_k$ as in \rf{3333}.
%
We will need the following small $\S$ expansions of the parameters:
\be
&&\kappa=\J+\frac{2+\J^2}{\J\sqrt{1+\J^2}}\S-\frac{2+6\J^2+3\J^4}{\J^3(1+\J^2)^2}\S^2+...\ , 
 \la{ex}
\\  &&
r=\frac{\S^{1/2}}{(1+\J^2)^{1/4}}-\frac{2+\J^2}{2(1+\J^2)^{7/4}}\S^{3/2}+...
\ , \ \ \ \ \nu=\J+\frac{\J\,\S}{\sqrt{1+\J^2}}+\frac{\J\;\S^2}{(1+\J^2)^2}+... \ . 
 \no
\ee
The corresponding characteristic polynomials are  
given in  Appendix B.3.
The summation prescription in \rf{BPT}  may be fixed as follows.
All frequencies which are nonzero in the BMN limit ($r\rightarrow 0$)
are summed with 
uniform signs such that at $p_1 \gg1 $ they contribute positively to the energy
(this guarantees the vanishing 
of 1-loop correction to the BMN vacuum state). The 
signs of some remaining frequencies
are fixed by requiring the absence of $r^2\ov \nu$ terms 
in the frequency sum. Few other signs 
are fixed by requiring that all $\OO(r^2)$ terms vanish
(such terms are  expected to cancel due to opposite curvatures of $AdS_5$ and $S^5$).
Then as in \rf{yyy},\rf{yy}  and \rf{er} we may split   the contribution 
of  modes with $p_1=-2,\dots,2$  from that of the higher ones
\be
E_{2d\ \rm low} &=& 
\Big[\frac{1}{2}\Big(-\frac{961}{72}-\frac{9}{8} u\Big)
                 +\frac{1}{2}\Big(\frac{141337}{10368}+
		 \frac{21}{16}u\Big)\nu^2+\OO(\nu^4)\Big]r^4+\OO(r^6) \ , \\
E_{2d\ \rm high}&=& 
 \sum_{p_1=3}^\infty \Big[\frac{ 4 p_1^4+11 p_1^2-3 }{p_1^3 (1-p_1^2)^3} 
+\OO(\nu^2)\Big]\ r^4+\OO(r^6)  
           \cr
           &=& 
	   \Big[\frac{1033}{144}-6\zeta_3+\OO(\nu^2)\Big]\ r^4+\OO(r^6) \ . \la{kky}
\ee
Here the parameter  $u$ represents the still unfixed sum of 4 bosonic $p_1 = \pm 2$ 
 frequency signs; it  can take values  $u=-4, -2, 0, 2, 4$.
Then $f_k$ in the analog of \rf{3333} are 
\be
f_0(\nu, 1)=0\ , \ \ \ \ \ 
\qquad
f_1(\nu, 1)=0\ , \ \ \ \ \ \ \ 
\qquad
f_2(\nu, 1) = \frac{1}{2} - \frac{9 u}{16} - 6 \zeta_3+\OO(\nu^2) \ . 
\ee
Expanding $E_1$ first in small $\S$  at fixed $\J$ and then in small $\J$ we  get 
\be
&&E_1= {1 \ov \kappa} E_{2d} = \Big[  \frac{2 n_{12}}{\J}
+\OO(\J)\Big] \S^2+\OO(\S^3) \ ,
\\
&& n_{12}=n'_{12}-3\zeta_3 \ , \ \ \ \ \ \ n'_{12}=\frac{8-9u}{32} \ .  \la{hhh}
\ee
This gives $n'_{12}= ( { 11 \ov 8}, \frac{13}{16}, \frac{1}{4})$   for $u=(-4, -2, 0)$.  
%
The choice of $n'_{12}=  { 11 \ov 8}$ appears to be preferred in the 
 algebraic curve approach that we discuss in Appendix C. 
 None of these choices leads to $n'_{12}= {5 \ov 8}$ consistent with 
the  universality of \rf{9} observed  for {\it four}
 other (two folded and two circular) 
examples of the solutions. 
This  suggests  that a consistent summation prescription in this $S=J'$ case is yet to be
identified. 

Expanding in $\S$  for fixed $\rho$  when  
\be
\nu = 2\rho{\S}^{1/2}+2\rho{\S}^{3/2}+\OO(\S^{5/2})
\ , \ \ \ \ \ \ 
\kappa=2\sqrt{1+\rho^2}{\S}^{1/2}-\frac{{\S}^{3/2}}{\sqrt{1+\rho^2}}+\OO(\S^{5/2}) \ , 
\ee
we get (cf. \rf{tuk},\rf{tik})
\be
E_1 = {\sqrt{\S}\ov \sqrt{1+\rho^2}} \ \Big[\ 
(n_{12}'-3\zeta_3 )\, \S +\OO(\S^2)\Big] \ , \la{typ}
\ee
where $n_{12}'$ is  the same as in  \rf{hhh}.

Similarly to the $J_1=J_2$ and $S_1=S_2$ cases in \rf{hightrJ1J2},\rf{hightrJ1J2x}, 
the transcendental parts of the higher terms 
in the small $\S$ expansion of $E_1$  here are found to be $(\N=2\S$)
\be
E_1 &=& \Big[
(\frac{n'_{12}}{2}-\frac{3}{2}\zeta_3)\frac{1}{\J}
 +(q_1+\frac{3}{2}\zeta_3+\frac{15}{8}\zeta_5)\J+\dots\Big]\N^2 \la{hightrJ1J2y}
\\
&&+\Big[\frac{q_2}{\J^5}+(q_3+\frac{3}{2} \zeta_3)\frac{1}{\J^3}+\frac{q_4}{\J}
       +(q_5-\frac{5}{8}\zeta_3-\frac{15}{16}\zeta_5-\frac{35}{16}\zeta_7)\J+\dots\Big]\N^3+
       \dots\no
\ee
where $q_{k}$ are rational numbers.
The coefficient of $\J \N^2$ term 
again leads to the same universal value of $\tn_{12}$ in \rf{155} with ${\tilde n}'_{12}=2q_1$. At $\OO(\N^3)$ 
we should find that $q_2=\frac{3}{4}n_{11}=0$ and that $q_3=-\frac{1}{2}n'_{12}$. 
The absence of $\zeta_5$ in $\N^3/\J$ term 
confirms again  the universality of $\zeta_5$ in  $n_{13}$ in \rf{uu}, the absence of $\zeta_3$ implies that 
$n_{13}''=  {3 \ov 2}$ and the rational term fixes $n'_{13}=2(q_1+q_4)+\frac{1}{2}n'_{12}$.

It is also possible to determine unambiguously 
the transcendental part of $E_1$ in the  small $\S$ expansion 
at $J=0$ (cf. \rf{ch},\rf{011})
\be
(E_1)_{\S=\J',\ \J=0}
=\sqrt{\S}\Big[ \ (n_{12}'-3\zeta_3) \S+
2(k_3+\frac{9}{4} \zeta_3+\frac{15}{4} \zeta_5) \S^2+\dots\Big] \ , 
\ee
where $k_3$ is a rational number.
 This again  leads  to $n_{13}$ in eq.~(\ref{uu}) 
with $n'_{13}=k_3+\frac{1}{4}n'_{12}$ and $n''_{13}= {3 \ov 2}$ (here $n_{02}=1$). Consistency of 
the two values for $n'_{13}$ requires then that $k_3=2(q_1+q_4)+\frac{1}{4}n'_{12}$.

\section*{Acknowledgments }

We would like to thank   B. Basso,  N. Gromov  and   A. Tirziu 
for    useful  discussions.  RR and AAT would like to thank Nordita for hospitality during part of this work while participating in the program on Exact Results in Gauge-String Dualities.
S.G. is supported by Perimeter Institute for Theoretical Physics. 
Research at Perimeter Institute is supported by the Government of Canada 
through Industry Canada and by the Province of Ontario through the Ministry 
of Research $\&$ Innovation.
The  work of RR  was supported by the US Department of Energy under contract
DE-FG02-201390ER40577 (OJI).
The work of AAT was supported by the ERC Advanced  grant No.290456.




\appendix
\section*{Appendix A: Comments on small  and large $\J$ 
expansions of   $h_1(\l,J) $ in eq. \rf{5} \la{A}  }

\refstepcounter{section}
\def\theequation{A.\arabic{equation}}
\setcounter{equation}{0}

Let us comment on the exact expression for the slope function $h_1(\l,J) $ in  \rf{5}
proposed in \cite{bas} in the case of the folded spinning string state in the $sl(2)$ sector 
and its possible generalizations for other string states. One motivation to try 
understand the
structure of $h_1$ better is that it determines, in particular,  the value of the 2-loop coefficient 
$n_{21}$ in \rf{3k} that is still to be derived by a direct world-sheet computation. 

It was  suggested  in \cite{bas}  that the exact  form of $h_1$
function in the energy (dimension)  \rf{5} of the    $sl(2)$ sector ground state 
corresponding in the  semiclassical limit to the $(S,J)$ 
folded string in $AdS_5$  is given by 
\be 
&& h_1= 
2 \sql {d \ov d \sql} \ln I_J(\sql)\la{4x}
 \\ &&  \ \ 
= 2 \sql \sqrt{ 1 + \J^2} -  { 1 \ov  1 + \J^2} -  {{1 \ov 4}  -\J^2  
 \ov \sql  (1+ \J^2)^{5/2} } -  { {1 \ov 4} - {5 \ov 2} \J^2 +  \J^4  
 \ov (\sql)^2  (1+ \J^2)^{4} }  + ...   \la{4a}\\  && 
  \ \ 
= 2 \sqrt{ \l  + J^2} -  { \l  \ov  \l + J^2} - { \l ({1 \ov 4} \l - J^2 ) \ov 
 (\l + J^2)^{5/2} }-   
 { \l ({1 \ov 4}\l^2 - {5 \ov 2} \l J^2 +  J^4  )
 \ov   (\l+ \J^2)^{4} }   +      ...\ , \la{4z}
\ee
where $I_J$ is the modified Bessel function and $\J = { J\ov \sql} $.
The second  line corresponds to the string semiclassical expansion: $\l \gg 1$
for fixed $\J$;  the first term in it is the classical string contribution, 
the second is 1-loop term, the third is 2-loop one, etc. 
The third line is found by  rewriting the semiclassical 
result back in terms of $J$.
 
 Starting with $E = \sqrt{ J^2 + h_1(\l, J)  N + ...}$ in \rf{5} 
 and expanding it in semiclassical regime  with 
  fixed $\J$  and small $\N$   we get 
  \be 
&&   E= J + { N \ov 2 J } h_1(\l, J)  + ... 
  = J + { N \ov 2 J } \big[ 2 \sqrt{ \l  + J^2} -  { \l \ov  \l + J^2} + ...\big]
 + ...
  \ .  \la{5552}
 \ee 
 The 1-loop term   $-{ \N \ov 2 \J }   { 1\ov  1 + \J^2}$
  was  found directly in the semiclassical  limit in  \cite{gva}.
  As was mentioned in section 1, this term is universal, i.e. 
  found also for other semiclassical states (see  \rf{666}).
 This expression   can be expanded in several different 
limits  and   interpolates between some   previously known results.  
If we  assume that $ \J \gg 1$, i.e.  $J^2 \gg \l$,   then we get from \rf{4a}
\be \la{5a}
h_1= 2 J + { \l \ov J} \Big(1 - { 1 \ov J}  + { 1 \ov J^2}+... \Big)  + ...\ , 
\ee
implying that the expansion of $E$  in the large $J$, small ${\N \ov \J}$ limit is 
\be \la{6a}
E= J + N + { \l \ov 2 J^2}  N \Big(1 - { 1 \ov J}   +  { 1 \ov J^2} + ... \Big) + 
O( ({N \ov J})^2)\ . 
\ee
This matches the known tree level plus 1-loop result in string semiclassical 
expansion.\foot{For folded string the $1\ov J$ term was found in Appendix D of \cite{mtt1}.}
Notice that in $(1 - { 1 \ov J}  + { 1 \ov J^2}+... )$ in \rf{5a} the string 
1-loop term $- { 1 \ov J}$ came from the $-  { \l  \ov  \l + J^2}$ term 
in \rf{4a}  while the  string 2-loop term
$ + { 1 \ov J^2}$   came from the $-{ \l ({1 \ov 4} \l -J^2  ) \ov 
 (\l + J^2)^{5/2} }$  term in \rf{4a}.
 
These two leading terms are, in fact, protected, i.e. are the same as on 
 the  1-loop  gauge theory (spin chain)  side  where the  $1 \ov J$ term is the leading 
finite size correction \cite{btz}. 
The structure $(1 - { 1 \ov J} )$ of the leading 
correction appears to be 
universal: it is found also for the circular $(S,J)$ 
string \cite{ptt,btz}.\foot{To see that
there is no linear in $S/J\equiv {N/J}$  term in the ``anomalous'' part of the 1-loop correction 
$E_{anom}= { \l \ov 2 J^2} ( \sum^\infty_{n=1}[  n  \sqrt{   n^2 + 4 M^2} - n^2 - 2 M^2]$
 where $M^2 = { S \ov J} (1 + { S \ov J}  )$  one needs to differentiate  this over $M^2$
(the first derivative vanishes).
 } 
 This is consistent with  the relations \rf{333},\rf{666}.
 The linear in
$N\ov J$ term   comes only from the  zero-mode   contribution on the string side or only from
the non-anomalous finite-size correction 
 on the 1-loop gauge theory side.
The next   $1 \ov J^2$ correction (1-loop on gauge theory side and 
  2-loop on the semiclassical string theory side) which should  again be protected 
 was computed  on the spin chain side in \cite{gk} (to all orders in $N\ov J$).\foot{As we 
 have checked explicitly from the results in Appendix of \cite{gk},  
the same subleading $1/J^2$ finite-size term as in \rf{6a}
appears also for the circular $(S,J)$ string state  in the $sl(2)$ sector 
(here $J$ is the momentum along the circle in $S^5$  which the string is wound on).
This suggests   the universality of  the terms given explicitly in \rf{6a}
in the $sl(2)$  sector.}

If instead we  consider the opposite limit of 
$ \J \ll 1$, i.e. $J^2 \ll \l$,    then we get from \rf{4} \cite{bas}
\be \la{7a}
h_1= 2 \sql - 1 - { {1  \ov 4} - J^2\ov  \sql }     - { {1  \ov 4} - J^2\ov  (\sql)^2 } 
- { { 25 \ov 64} - {13\ov 8} J^2 + {1  \ov 4}  J^4 \ov (\sql)^3} + 
  ...\ , 
\ee
implying the values $n_{11}=-1, \ \nm=1$ and
 $n_{21}=-{1 \ov 4}$ in \rf{3k},\rf{12}.
This value  for the 1-loop  coefficient $n_{11}$
in the small $\S$ semiclassical expansion 
(matching the one directly computed 
using the algebraic curve approach in \cite{gssv})  
 has the same origin in \rf{4a} as the 
 $1 \ov J$ string term in \rf{6a}: both come from two different limits of the 
 the 
1-loop semiclassical term 
$-  { 1  \ov  1 + \J^2} $  there. 
This  confirms that this term should not   be
sensitive to wrapping (Luscher) corrections,    being at the same time 
the origin of a finite-size
(and even non-anomalous)  term  at large $J$.
 This also suggests that, like the coefficient of the  $- {1 \ov
J}$ term,  $n_{11}$  may be coming only from the zero-mode contributions  in the 
near folded-string expansion.  This  supports the claim  \cite{bas} 
that $h_1 (\l, J)$ has its origin just  in the 
asymptotic Bethe ansatz   and is not even sensitive to the string phase.

One may expect to find similar expressions for the corresponding $(J', J)$  
folded string 
state in the $su(2)$ sector. Indeed, the  folded string in $S^5$ is related 
to  its $AdS_5$ counterpart by  an analytic continuation \cite{bfst}, 
 implying (up to signs)
$(E, S; J) \to (E; J', J), \   E= -J, \ S=J', \ J =- E$.
In this case $N= J'$  so  we may  expect to  get 
similar  relations as above up to some
 sign changes, i.e.\foot{Note  that this analytic continuation is not useful 
if  $J$ is fixed, while   $E \sim  \l^{1/4 }\gg  1$ 
so  there is no way of interchanging   $E$ and $J$.
It still  works   at large $\J$ and thus large $\E$
and explains why the  sign of first  finite size correction changes:
$E= J + { \l N \ov 2 J^2} (1 - J^{-1} + J^{-2}) $  translates into  
$J= E - { \l N \ov 2 E^2} (1 + E^{-1} + E^{-2}) $  and then  using that 
$E=J + ...$  we get the required result.}  
\be \la{9a}
E^2= J^2 +  h_1(\l, J)\  J'   + ...\ ,  \
\ \ \ \ \ \ \ \ \  h_1  = 2\sql  \sqrt{ 1  + \J^2}  +  { 1  \ov  1 + \J^2} + ...\ . 
\ee
Changing the  of sign of the subleading term in \rf{9a} compared to \rf{4a} 
has two implications: the  signs of 
$n_{11}$, of $\nm$   and of the  leading $1 \ov J$ term also change.  
Now  $n_{11} =  { 1 }=- \nm $ as in   \rf{13}  in agreement  
with  
\cite{rt1,bm} (see also Appendix D).\foot{The change of sign of the leading 1-loop
 string correction can be attributed to the 
change in sign of the curvature  between $AdS_5$ and $S^5$  \cite{rt1}.} 
For large $\J$ we get 
\be \la{10a}
E= J + J' + { \l \ov 2 J^2}  J' (1 + { 1 \ov J}  + { 1\ov J^2} +... ) +  ...\ , 
\ee
where the  $(1 + { 1 \ov J}  )$ term  is in agreement with  the 
result for the finite size corrections from the spin chain and the string sides
(cf. eq.7.33,7.34  in \cite{mtt1}).
As in the  $sl(2)$ sector 
 case in \rf{4a},\rf{5a}, the 
 subleading term ${ 1\ov J^2}$  in \rf{10a} 
 should  originate from the next (string 2-loop) 
 term in  $h_1$ in \rf{9a}. 
 %
%
The coefficient of this ${ 1\ov J^2}$ term  
should be universal in the $su(2)$ sector, i.e. the 
same also as for the circular  string. 
Indeed,  for the circular  string in the $su(2)$  sector
we get \rf{10a}   with the same terms in the 
bracket  $( 1 +  {1\ov J} + {1\ov J^2} + ... )$, as 
one can see from  
 \cite{grig}  (these terms   come from  non-anomalous finite size contribution only).
 Such  a correction in the near-BMN expansion was found also 
 in \cite{mtt1}. It came out  the  same from the Bethe ansatz
   and the Landau-Lifshitz approach,  
so it should be a protected one.\foot{The fact that it comes out of 
the  Landau-Lifshitz approach means that one does  not need the 
full superstring computation to reproduce it,  provided one regularizes properly
(in addition, only zero modes are expected to contribute to this term).} 
Direct check of the universality of the ${1\ov J^2}$ term 
  requires a 2-loop computation on the  string side. 
The knowledge  of this ${1\ov J^2}$ term   provides a priori 
only  a weak constraint on a possible 
next term in the expansion of $h_1$ in \rf{9a}, but there is a natural guess: 
  the direct analog of
the $ - { \l ({1 \ov 4} \l- J^2  ) \ov  (\l + J^2)^{5/2} }$ 
term in \rf{4z}   reproduces both the ${1\ov J^2}$ term
 and the expected  universal 
value of $n_{21}$ in \rf{9} (see \rf{13}).


In   the case of ``small''  circular strings with 2 internal spins
we again find 
\be \la{21}
 h_1 = 2\sql \sqrt{1 + \J^2 }  +   { n_{11}    \ov 1 + \J^2 } +   ... \  ,   
\ee
where 
e.g., for $(J_1=J_2=J', J)$ case  $N=J_1+J_2=2J'$ and 
 $n_{11}=2=-\nm$  (see \rf{14}).
 Indeed, according to \rf{fe}, in this case 
\be 
E_1 = { \N \ov \J (1 + \J^2)}  + O(\N^2)   \ , \ \ \ \ \ \ \ \ \ 
 \la{22}
h_1(J \gg \sql ) = 2 J \Big[ 1  + { \l \ov 2 J^2} ( 1 + { n_{11} \ov   J})
 + ... \Big] \ .  
\ee
The  term   $ 1 + { n_{11} \ov   J}$ with $n_{11}=2$ 
 here  appears to be  in contradiction 
 with   the form of the  finite size correction --  
$ (1 + { 1 \ov J})$ times  the  classical $ {\l \ov 2  J^2} N $  term --
found earlier  \cite{fk1,btz}.\foot{This structure 
 from expansion of eq.(2.23) in \cite{btz} 
to linear order
in $\N$: again only the 
analytic spin chain side part or 0-mode string side  part is  contributing
to it. 
It appears that the analytic finite size correction to the linear  in $\N$ term 
is universal:
$1 + {1 \ov L} $   in compact  (su(2), etc.)
 sector and  $1 - {1 \ov L}$ in noncompact (sl(2), etc)
 sector. Here   $L= J + N $ is total 
 length, its difference from $J$ is irrelevant to leading order in $N$. The sign
 difference is due to the analytic continuation between the sectors.}
 As already mentioned below eq. \rf{26} this  is not really a disagreement 
 as,  in the {\it 2-spin case},  the two expressions are derived in different limits: here we have $\J' \ll 1$ for fixed $\J$, while 
 in the standard discussions of finite-size corrections in the thermodynamic limit one first assumes $ \J' \gg 1, \ \J \gg 1$, with $ \J' \ov \J$=fixed, 
   and then may expand  in $ \J' \ov \J$. 

\section*{Appendix B: Characteristic polynomials for circular string\\  fluctuation
frequencies   \label{App:J1eqJ2}  }

\refstepcounter{section}
\def\theequation{B.\arabic{equation}}
\setcounter{equation}{0}

Rigid circular strings with two equal spins  and orbital momentum $J$ in $S^5$ discussed in
this paper  are homogeneous  solutions  for which the quadratic fluctuation operator has
constant coefficients. In Fourier transformed form this is a matrix depending 
on 2d momenta  $(p_0,p_1)$  (with $p_1$ being integer as $\sigma \in (0, 2\pi)$) 
whose  determinant is thus a finite-order polynomial in  $(p_0,p_1)$.
The roots of this characteristic polynomial 
determine the fluctuation frequencies 
$p_0=\omega(p_1)$ that appear in the 1-loop  correction to 2d energy 
(see \rf{2.1} or \rf{BPT}). While we focused on the solutions with unit winding number, $m=1$, 
a nontrivial value of $m$ may be introduced in the characteristic equations for all three circular 
string solutions through the formal rescaling,
\be
&&
p_0\rightarrow \frac{p_0}{m}
\quad
p_1\rightarrow \frac{p_1}{m}
\quad
\kappa\rightarrow \frac{\kappa}{m}
\quad
\nu\rightarrow \frac{\nu}{m}
\quad
w\rightarrow \frac{w}{m}
\quad
w'\rightarrow \frac{w'}{m}
\quad
r\rightarrow r
\quad
a\rightarrow a \ .
\ee
This rescaling may be identified in the classical solutions \rf{hh}, \rf{j}, and \rf{mm}.

\subsection{ $J_1=J_2$ string }

The   characteristic polynomials for this  circular string have been derived in 
\cite{ft2,ft3}.
The $AdS_5$ fluctuations have the standard BMN type form with mass  $\kappa$ 
(expressed in terms of the other independent parameters  $a$ and $\nu$, see \rf{hh})
while the characteristic polynomial for the   $S^5$ part is more complicated. Explicitly, 
\be
B^{AdS_5}_8&=&\big(-p_0^2 + p_1^2 + \nu^2+4 m^2 a^2\big)^4\ , 
\\[3pt]
B^{S^5}_8&=&\big[(p_0^2-p_1^2)^2-4 \nu ^2 p_0^2\big]^2
-16 (2 a^2-1) m^4 (p_0^2-p_1^2)^2
\\
& & + 8 m^2 \Big[(a^2-1) (p_0^2-p_1^2)^2
   (p_0^2+p_1^2)-4 \nu ^2 p_0^2 [(a^2-1) p_0^2+(1-3 a^2)
   p_1^2]\Big]
\nonumber
\end{eqnarray}
As discussed in \cite{ft2, ft3}, the determinant of the fermionic quadratic operator is the square
of an operator expressed solely in terms of six-dimensional Dirac matrices. We note here that, due 
to the chirality of six-dimensional spinors, this determinant (over spinor indices)  further factorizes: 
\be
\det K_f^{10d}=(\det K_f^{6d})^2 \ , \ \ \ \ 
\qquad
\det K_f^{6d} = F_1F_2 \ , \la{ppp}
\ee       
where $F_{1,2}$ are the corresponding fermionic characteristic polynomials
\be       
F_1 &=& (p_0^2-p_1^2)^2  
+p_0^2 [\nu  (-\sqrt{4 a^2 {m}^2+\nu ^2}- 3 \nu )-2 (a^2+1) {m}^2]\no
\\
& &+ p_1^2 [\nu  (\nu  -\sqrt{4 a^2 {m}^2+\nu ^2})+(6 a^2-2) {m}^2]
\\
& &+(a^2-1)^2 {m}^4 + {m}^2 \nu  [\nu+ (a^2-1) \sqrt{4 a^2 {m}^2+\nu ^2}]+\ha \nu ^3
   (\nu -\sqrt{4 a^2 {m}^2+\nu ^2})   
   \ , \no
\\
F_2 &=&(p_0^2-p_1^2)^2
+p_0^2 [\nu  (\sqrt{4 a^2 {m}^2+\nu ^2}-3 \nu )-2 (a^2+1) {m}^2]\no
\\
& &+ p_1^2 [\nu  ( \nu + \sqrt{4 a^2 {m}^2+\nu ^2}
   )+(6 a^2-2) {m}^2]
\\
& & +(a^2-1)^2 {m}^4+ {m}^2 \nu  [\nu -(a^2-1) \sqrt{4 a^2 {m}^2+\nu
   ^2}]+   \ha \nu ^3 (\nu + \sqrt{4 a^2 {m}^2+\nu ^2} )  \ .
\no
\ee       
Using  the relations between the parameters of the solution, one can check that 
the product $F_1F_2$ reproduces the fermionic  characteristic polynomial in \cite{ft3}.

\subsection{ $S_1=S_2$ string }

As was mentioned in section 2, this solution may be obtained from the  $J_1=J_2, J$ by the analytic continuation
\be 
\kappa\leftrightarrow \nu\ 
, \  \ \ \ \ \ \ \ \ 
a^2\leftrightarrow -r^2 \  .
\ee
This observation may be used to find the  corresponding characteristic polynomials 
from their $J_1=J_2$ counterparts. The bosonic  ones  are then
\be
B^{AdS_5}_8&=&\big[(p_0^2-p_1^2)^2-4 \kappa^2 p_0^2\big]^2
-16 (2 r^2-1) m^4 (p_0^2-p_1^2)^2\no 
\\
& & +8 m^2 \Big[(r^2-1) (p_0^2-p_1^2)^2
   (p_0^2+p_1^2)-4 \kappa ^2 p_0^2 [(r^2-1) p_0^2+(1-3 r^2)
   p_1^2]\Big]
\  , 
\\
B^{S^5}_8&=&\big(-p_0^2 + p_1^2 + \nu^2\big)^4\ . 
\end{eqnarray}
The fermionic  determinant  has factorization property similar to that in 
 the $J_1=J_2,J$ solution  \rf{ppp} with
\be
&&F_1 = (p_0^2-p_1^2)^2  
+p_0^2 [-\kappa  (\nu+3 \kappa )-2 (-r^2+1) {m}^2]\no
\\
&&\ \ \  +p_1^2 [\kappa  (\kappa -\nu) -2(3 r^2+1) {m}^2]
\no\\
&&\ \ \  +(r^2+1)^2 {m}^4 +  {m}^2 \kappa  [\k- (r^2+1) \nu  ]   +\ha \kappa^3(\kappa -\nu)
 \ , 
\\
&&F_2 =(p_0^2-p_1^2)^2
+p_0^2 [\kappa  (\nu-3 \kappa )-2 (-r^2+1) {m}^2]\no
\\
&&\ \ \  +p_1^2 [\nu  (\nu+\kappa) -2(3 r^2+1)   {m}^2]
\cr
&&\ \ \ + (r^2+1)^2 {m}^4+ {m}^2 \kappa  [\kappa + (r^2+1) \nu]+\ha \kappa^3 (\k + \nu)
\ .
\ee       
Upon setting $\nu=0$ we may recover the characteristic 
polynomials in \cite{PTT}.

\subsection{$S=J'$ string}

Here the $AdS_5$ bosonic characteristic polynomial  can be directly  extracted  from \cite{ptt}
(from the expression found before using the conformal gauge constraint).\foot{One can check 
directly that the massless mode decouples in the characteristic polynomial 
for three coupled $AdS_5$  fluctuation modes
that follows from the fluctuation Lagrangian in eq. (4.13) in \cite{ptt}.}
Then its $S^5$ counterpart  can be found by using the ``self-duality'' property of the solution 
\rf{mm} under
  \be \k \leftrightarrow \nu\ ,\ \ \ \ \  \ r\leftrightarrow i a\ 
,\ \ \ \ \  \ w \leftrightarrow - w' \ . \ee
Including also a nontrivial winding number $m$, we end up with 
\be
\label{Ad}
&& B^{AdS_5}_8= 
( -  p_0^2 +  p_1^2 + w^2-m^2)^2 \cr
&&
 \ \  \times\Big[ (p_0^2-p_1^2)^2 - 4 m^2p_1^2(1 + r^2) + 
   8  mp_0  p_1 (1 + r^2) w 
  - 4  p_0^2 [-\kappa^2 r^2 + (1 + r^2) w^2]\Big] \\
&&
B^{S^5}_8=
( -  p_0^2 +  p_1^2 + w'^2-m^2)^2 \cr
&&
\ \ \times\Big[ (p_0^2-p_1^2)^2  -4 m^2 p_1^2(1 - a^2)   - 8 m p_0  p_1  (1 - a^2) w' - 
   4  p_0^2 [\nu^2 a^2 + (1 - a^2) w'^2]\Big]
   \ . ~~~~\label{sc}
\ee
As in the previous cases  here the fermionic operator can be put into a 
%
%
block-diagonal form where  each block may be written 
in terms of the six-dimensional Dirac matrices. While the two blocks are not identical, parity invariance requires 
that their determinants are the same.
The fact that six-dimensional spinors are chiral implies
 that the determinant of each block further factorizes as in \rf{ppp}, where now 
\be
F_1 &=& (p_0^2-p_1^2)^2+2 m  p_0   p_1 \Big[2 a^2 (w'{}+\frac{\kappa  \nu}{w} )+(w-w'{})\Big]
   \cr
   &&+ p_1^2  \Big[-\kappa  \nu +3 \nu ^2+(w-2 w'{})
   (w+w'{})\Big]-p_0 ^2 \Big[\kappa\nu +\nu^2 +w (w+w'{})\Big]\la{jjb}
  \\
   &&+\frac{1}{4}  \Big[-2 \kappa  \nu  [w'{}
   (w+w'{})-\nu ^2]+2 \nu ^4+\nu ^2 (w-3 w'{}) (w+w'{})+w'{}^2
   (w+w'{})^2\Big] \ , \no 
\\
F_2&=&(p_0^2-p_1^2)^2+2 m  p_0   p_1 \Big[2a^2 ( w'{}- \frac{\kappa  \nu}{w} )+(w-w'{})\Big]
   \cr
   &&+ p_1^2 \Big[\kappa\nu +3 \nu^2 +(w-2 w'{}) (w+w'{})\Big]- p_0 ^2  \Big[-\kappa  \nu +\nu ^2+w
   (w+w'{})\Big] \la{jja}
   \\
   &&+\frac{1}{4} \Big[2 \kappa  \nu  [w'{} (w+w'{})-\nu
   ^2] +2 \nu ^4+\nu ^2 (w-3 w'{}) (w+w'{})+w'{}^2
   (w+w'{})^2\Big] \ .\no
\ee
Let us comment on derivation of these expressions (that reduce to the ones in \cite{ptt} for
$a=1$ in \rf{mm}). 
In  the $\k$-symmetry  gauge $\theta_1=\theta_2$ the quadratic part of the 
 fermionic Lagrangian is (see, e.g., \cite{ft3,ptt} and refs. there)
\be
L = -2i\eta^{\alpha\beta} e_\alpha^A{\bar\theta}\Gamma^A{\cal D}_\beta\theta
   -\epsilon^{\alpha\beta}{\bar\theta}\Gamma_A\Gamma_*\Gamma_B\theta
e_\alpha^A e_\beta^B
\label{Lf}
\ee
where ${\cal D}=d+\frac{1}{4}\omega^{AB}\Gamma_{AB} $ is the usual spinor covariant derivative.
For the solution \rf{mm} the 2d projected combinations 
 $e_\alpha^A\Gamma_A$ and  $\omega_\alpha^{AB}\Gamma_{AB}$ are:
\be
e_0^A\Gamma_A&=&\Gamma_0 \sqrt{1 + r^2} \kappa + \Gamma_4 r w 
+ \Gamma_5 \sqrt{1 - a^2}\nu + \Gamma_9 a w' \no
\\
e_1^A\Gamma_A&=&m (\Gamma_4 r - \Gamma_9 a)\no
\\
\omega_0^{AB}\Gamma_{AB}&=&
2 \kappa r \Gamma_{01} - 2 (\sqrt{1 + r^2} w \Gamma_{14} + a \nu \Gamma_{5 6} 
+  \sqrt{1 - a^2} w' \Gamma_{69})\no
\\
\omega_1^{AB}\Gamma_{AB}&=&
m (-2 \sqrt{1 + r^2} \Gamma_{1 4} + 2 \sqrt{1 - a^2} \Gamma_{6 9})
\label{parts}
\ee
where $\Gamma_A$ are the 10-d Dirac matrices;  one should project the quadratic operator
onto its chiral part  thus rendering it a $16\times 16$ matrix.
To evaluate the determinant of the quadratic fermionic 
 operator we first  notice that the
matrices $\Gamma_{2}$ and $\Gamma_{3}$   in 
$\Gamma_* = i\Gamma_{01234}$ in \rf{Lf} do not appear elsewhere 
in the quadratic operator. The product 
$\Gamma_{23}$ may therefore be diagonalized; its diagonal entries are $\pm i$.
In this representation the quadratic 
operator is block-diagonal and each block may be obtained 
from (\ref{Lf}) and (\ref{parts}) by using for
 $\Gamma_i$ and $\Gamma_{ij}$ the d=6 Dirac matrices 
and $\Gamma_* = \pm \Gamma_{014}$. Since the sign of $\Gamma_*$ affects only the sign of the 
Wess-Zumino term which can also be changed by parity transformations, the determinants of the two blocks
are equal and thus the 10d determinant is a perfect square, as in  the
 first equation in (\ref{ppp}).
Since the 6d spinors are chiral, there exists a 
representation of the 6d Dirac matrices in which each block 
of the quadratic operator is itself block-diagonal. 
Thus, the determinant of each block further factorizes; 
each block is only a $4\times 4$ matrix and its determinant can be easily evaluated
leading to the  two factors 
$ F_1$ and $F_2$ in eq.~(\ref{ppp}) given by \rf{jjb},\rf{jja}.

In section ~\ref{SeqJp} we discussed the small $r$ expansion
 of the energy of the $S=J'$ string 
with angular momentum $J$. For this purpose we need that
\be
a&=&r\sqrt{1+\frac{2r^2}{1+\nu^2}}
\ , \qquad
\kappa=\sqrt{\nu^2+4r^2+\frac{4r^4}{1+\nu^2}}\ , 
\\
w &=&\sqrt{1+\nu^2+4r^2+\frac{4r^4}{1+\nu^2}}\ , 
\qquad
w'=\sqrt{1+\nu^2}\ . 
\ee
Plugging these expressions in $F_1$ and $F_2$ and dividing by a factor of $r^4$ we find that
\be
F_{1,2} =c^{(1,2)}_0+c^{(1,2)}_2r^2+c^{(1,2)}_4r^4+...  \ , 
\ee
with
\be
c^{(1)}_0&=&c^{(2)}_0=
\left(\nu ^2-p_0^2+p_1^2-2p_1+1\right) \left(\nu ^2-p_0^2+p_1^2+2p_1+1\right)
\cr
c^{(1)}_2 &=& \frac{8}{(1+\nu^2)^{3/2}}\Big[
\sqrt{\nu ^2+1} \left(-2p_0^2 \left(4 \nu ^2+p_1^2+3\right)+p_0^4
+\left(p_1^2-1\right)^2\right)
\cr
&&
\qquad\qquad
+4
   \left(\nu ^2+1\right)^2p_0p_1\Big]
\\
c^{(2)}_2&=&\frac{8}{(1+\nu^2)^{3/2}}\Big[
4 \left(\nu ^2+1\right)p_0p_1+\sqrt{\nu ^2+1} \big(3 \nu ^4+\nu ^2 \left(-4p_0^2+4
  p_1^2+6\right)
   \cr
&&
\qquad\qquad   
+p_0^4-2p_0^2 \left(p_1^2+2\right)+p_1^4+3\big)
   \Big]
\\
c^{(1)}_4&=&\frac{4}{\nu^2(1+\nu^2)^{5/2}}
\Big[
32 \left(\nu ^3+\nu \right)^2p_0{}p_1{}+\sqrt{\nu ^2+1} \big(4 \nu ^4 \left(p_1^2-6p_0^2\right)
\cr
&&
\qquad\qquad   
+\nu ^2
   \left(p_0^4-2p_0^2 \left(p_1^2+10\right)+p_1^4+4p_1^2+3\right)+2
   \left(p_0^2+p_1^2+1\right)\big)
\Big]
\\
c^{(2)}_4&=&\frac{4}{\nu^2(1+\nu^2)^{5/2}}
\Big[
16 \left(\nu ^2+1\right) \nu ^2p_0p_1+\sqrt{\nu ^2+1} \big(13 \nu ^6+\nu ^4 \left(-14p_0^2+14
  p_1^2+24\right)
\cr   
&&
\qquad\qquad   
   +\nu ^2 \left(p_0^4-2p_0^2 \left(p_1^2+8\right)+p_1^4+8p_1^2+9\right)-2
   \left(p_0^2+p_1^2+1\right)\big)
\Big]
\ee
It is not difficult to construct higher orders in the small $r$ expansion at fixed $\nu$.


\section*{Appendix C: One-loop energy of  $S=J'$ circular string from the\\  algebraic curve
approach}

\refstepcounter{section}
\def\theequation{C.\arabic{equation}}
\setcounter{equation}{0}

Here we shall revisit the computation of the 1-loop correction to the energy of the 
$S=J'$ circular string 
solution \rf{mm} discussed in section 2.4  using the algebraic curve approach \cite{gv2,Gromov:2008ec}
 to determine the
fluctuation frequencies. 

In order to focus on  a near flat space expansion in the short string limit we will 
consider the limit $\S\to 0$ for fixed   $\rrho$ 
\begin{equation}
 \rrho=  { \nu \ov  2\,\sqrt{\mathcal{S}}} \ .   \la{rr}
\end{equation}
In section 2.4 in \rf{rre} 
we used  instead 
\be 
\rho= { \J \ov 2 \sqrt \S} =   \rrho \Big( 1 
- \frac{ \mathcal{S}}{\sqrt{1 + 4 \rrho^2 \mathcal{S}}} \Big).
\ee
Note also that 
\begin{equation}
  \mathcal{S} = \frac{\oomega^2 \sqrt{1 + 2 \rrho^2 \oomega + \rrho^4 \oomega^4} - 
\omega (1 + 2 \rrho^2 \oomega - \rrho^2 \oomega^3) }{2 (1 + 2 \rrho^2 \oomega)^2}\ .
\end{equation}
%


\subsection{Quasimomenta}

The quasimomenta can be obtained 
by explicit diagonalization of the monodromy matrix~\cite{gv2}; for the $S^5$ part the basic 
single cut quasimomenta  vanishing at infinity  are determined by 
\begin{equation}
  \tilde{p}(x) = -\pi  + \pi  \frac{x - \tilde{x}_1}{x^2 - 1} 
    \sqrt{(x - \tilde{x}_2)(x - \bar{\tilde{x}}_2)} \  ,
\end{equation} 
where the two roots $\tilde{x}_1, \tilde{x}_2$ are given by:
\begin{eqnarray}
  && \tilde{x}_1 = - \frac{1}{2 \rrho \sqrt{\mathcal{S}}
   + \sqrt{1 + 4 \rrho^2 \mathcal{S} }}\ , \nonumber\\
  && \tilde{x}_2 = \frac{
    \big( \sqrt{1 + 4 \rrho^2 \mathcal{S} } + 2 \rrho 
    \sqrt{\mathcal{S}} \big)\Big( \sqrt{1 + 4 \rrho^2 \mathcal{S} } + 2 i 
      \sqrt{\mathcal{S} \big( \sqrt{1 + 4 \rrho^2 \mathcal{S} } -
       \mathcal{S}  \big)} - 2 \mathcal{S} \Big)}{\sqrt{1 + 4 \rrho^2 \mathcal{S} }}.
\end{eqnarray} 
The four $S^5$ quasimomenta can be identified looking at the 
asymptotic $x \to \infty$ behaviour of $\tilde{p}(x)$ and  $\tilde{p}({x^{-1}})$, which is related 
to the conserved global charges:
\begin{eqnarray}
  \frac{x}{2 \pi}  \tilde{p}(x) \to  \mathcal{S} - \mathcal{J} + \dots\ , \ \ \ \ \ 
  \frac{x}{2 \pi}  \tilde{p}({x^{-1}}) \to -1 - \mathcal{S} - \mathcal{J} + \dots
\end{eqnarray}
Hence, we can identify:
\begin{eqnarray}
  p_{\tilde{1}}(x)= - 2 \pi - \tilde{p}({x^{-1}})\ , \ \ \ \
  p_{\tilde{2}}(x)= \tilde{p}(x)\ , \ \ \ \
 p_{\tilde{3}}(x)= - \tilde{p}_2(x)\ , \ \ \ \ 
  p_{\tilde{4}}(x)= - \tilde{p}_1(x).
\end{eqnarray}
For the $AdS_5$ quasimomenta, the basic function is given by: 
\begin{equation}
  \hat{p}(x)=\pi  \frac{x -\hat{x}_3}{x^2-1} \left( \sqrt{x 
  - \hat{x}_1}\sqrt{x - \hat{x}_2} -1 \right) \ ,
\end{equation}
where the $\hat{x}_i$ are:
\begin{eqnarray}&&\hat{x}_1 = \left( \hat{x}_2 \hat{x}_3^2 \right)^{-1} \ , \ \ \ \ \ 
  \hat{x}_2 = - \frac{2 \mathcal{S} + \oomega - 
  2 \sqrt{\mathcal{S}(\mathcal{S} + \oomega)}}{\oomega(\oomega - \sqrt{\oomega^2-1})}
  \ , \ \ \ \ \ 
  \hat{x}_3 = \oomega - \sqrt{\oomega^2 - 1} \ . 
\end{eqnarray}
Again, comparing with the asymptotic, the identification of the quasimomenta goes as follows:
\begin{eqnarray}
  p_{\hat{1}}(x)= - \hat{p}({x^{-1}})\  , \ \ \ \ 
  p_{\hat{2}}(x)= \hat{p}(x)\ , \ \ \ \ 
  p_{\hat{3}}(x)= - \hat{p}(x)\ , \ \ \ \ 
  p_{\hat{4}}(x)= \hat{p}({x^{-1}}) \ .
\end{eqnarray}


\subsection{Off-shell frequencies}

Due to the symmetry of the circular string solution, all the fluctuation energies
can be conveniently written as combinations of only two 
independent functions $\Omega_A(x) = \Omega^{\hat{2}\hat{3}}(x)$ and 
$\Omega_S(x) = \Omega^{\tilde{2}\tilde{3}}(x) $~\cite{Gromov:2008ec}:
\begin{eqnarray}
  &&\Omega_{B_1}(x) =  \Omega^{\tilde{1}\tilde{4}}(x) = - \Omega_S({x^{-1}}) + \Omega_S(0) \nonumber\\
  &&\Omega_{B_2}(x) =  \Omega^{\tilde{2}\tilde{4}}(x) = \Omega^{\tilde{1}\tilde{3}}(x) = \frac{1}{2}\left[\Omega_S(x) - \Omega_S({x^{-1}}) + \Omega_S(0)  \right]\nonumber\\
  &&\Omega_{B_3}(x) =  \Omega^{\hat{1}\hat{4}}(x) =  - \Omega_A({x^{-1}}) - 2  \nonumber\\
  &&\Omega_{B_4}(x) =  \Omega^{\hat{2}\hat{4}}(x) = \Omega^{\hat{1}\hat{3}}(x) = \frac{1}{2}\left[\Omega_A(x) - \Omega_A({x^{-1}}) \right] - 1\nonumber\\
  &&\Omega_{F_1}(x) =  \Omega^{\hat{2}\tilde{4}}(x) =  \Omega^{\tilde{1}\hat{3}}(x) =  \frac{1}{2}\left[\Omega_A(x) - \Omega_S({x^{-1}}) +  \Omega_S(0) \right]\nonumber\\
  &&\Omega_{F_2}(x) =  \Omega^{\tilde{2}\hat{4}}(x) =  \Omega^{\hat{1}\tilde{3}}(x) =  \frac{1}{2}\left[\Omega_S(x) - \Omega_A({x^{-1}}) \right] - 1\nonumber\\
  &&\Omega_{F_3}(x) =  \Omega^{\tilde{1}\hat{4}}(x) =  \Omega^{\hat{1}\tilde{4}}(x) =  \frac{1}{2}\left[-\Omega_S(x) - \Omega_A({x^{-1}}) +  \Omega_S(0) \right] - 1\nonumber\\
  &&\Omega_{F_4}(x) =  \Omega^{\hat{2}\tilde{3}}(x) =  \Omega^{\tilde{2}\hat{3}}(x) =  \frac{1}{2}\left[\Omega_A(x) - \Omega_A(x) \right].
\end{eqnarray}
Following \cite{Gromov:2008ec}, the two functions $\Omega_A(x)$ and $\Omega_S(x)$ can be uniquely fixed 
imposing the correct analytical and asymptotic 
properties for the perturbed quasimomenta $p + \delta p$: 
\begin{eqnarray}
  &&\Omega_S(x) =  \Omega^{\tilde{2}\tilde{3}}(x) = \frac{\hat{f}(1)}{\tilde{f}(1)}\Big( 
  \frac{\tilde{f}(x)}{x-1} -1  \Big) + 
  \frac{\hat{f}(-1)}{\tilde{f}(-1)}\Big( \frac{\tilde{f}(x)}{x+1} -1  \Big) \ , \nonumber\\
  && \Omega_A(x) =\Omega^{\hat{2}\hat{3}}(x) = 2 \left( \frac{x}{x^2-1} \hat{f}(x) -1 \right),
\end{eqnarray}
where the two functions $\hat{f}(x)$ and  $\tilde{f}(x)$ are defined as
\begin{eqnarray}
  \tilde{f}(x) = \sqrt{(x - \tilde{x}_2)(x - \bar{\tilde{x}}_2)}\ , \ \ \
  \ \ \ 
  \hat{f}(x) =\sqrt{(x - \hat{x}_1)(x - \hat{x}_2)}\ ,
\end{eqnarray}
with a suitable choice of the cuts.


\subsection{One-loop energy}

Given the above set of off-shell frequencies $\Omega_{I} = \Omega^{i,j}$, $I \in
 \{ A,S,B_{1,2,3,4},F_{1,2,3,4} \} $,  
the corresponding physical on-shell fluctuations energies 
associated to the $(i,j)$ excitations with mode number $n$, are given by 
\begin{equation}
\omega_{I}^{(n)} = \omega^{(n)}_{i,j} = \Omega^{i,j}(x^{i,j}_n)\ ,
\end{equation}
where, for any pair $(i,j)$, $x^{i,j}_n$ is determined as the solution of the equation 
\begin{equation}
p_i(x^{i,j}_n) - p_j(x^{i,j}_n) = 2 \pi n\ .
\end{equation}
The one-loop  correction to the energy can be obtained as a sum over $n$ and polarizations~\footnote{
In the algebraic curve formalism, the on-shell energies $\omega_{i,j}^{(n)}$ enter directly $E_{1}$ and do not require $1/\kappa$ factors.}
\begin{equation} 
E_1 = \frac{1}{2} \sum_{n = -\infty}^{+ \infty} \sum_{i,j} (-1)^{F_{i,j}}  \omega_{i,j}^{(n)}.
\end{equation}
This sum is sensitive to integer shifts in the labeling of the frequencies $n \to n+\delta$;
following \cite{gv2} here we propose to use the following choice:
\begin{eqnarray}
  E_1 = \frac{1}{2} \sum_{n = -\infty}^{+ \infty } \left[ 
      \omega_{S}^{(n-1)} + 
      \omega_{A}^{(n-1)} + 
      \omega_{B_1}^{(n-1)} + 
      \omega_{B_2}^{(n-1)} + 
      \omega_{B_3}^{(n+1)} + 
      \omega_{B_4}^{(n)}  \right.  \nonumber\\ 
      \left. - 2 \omega_{F_1}^{(n-1)} - 
      2 \omega_{F_2}^{(n)} -
      2 \omega_{F_3}^{(n)} - 
      2 \omega_{F_4}^{(n-1)} \right] \ . \la{pre}
\end{eqnarray}
Then  the final result in the short string limit has the same form 
 as in  \rf{typ}) 
\begin{equation}
  E_1 = \frac{{11\ov 8} - 3 \zeta_3}{  \sqrt{\rrho^2 + 1}}\,  \mathcal{S}^{3/2}\  + \mathcal{O}(\mathcal{S}^2) \ , 
\la{e11}
\end{equation}
corresponding to the rational part of $n_{12}$ in \rf{8},\rf{hhh}  being 
\be n'_{12} = {11 \ov 8}  \ .  \la{fia} \ee 
The prescription \rf{pre}  thus does not lead  to the preferred choice 
$n'_{12} = {5 \ov 8} $ consistent with the universal  value \rf{9} of the 2-loop coefficient $n_{21}$.
The  value in \rf{fia} together  with universality of Konishi dimension 
implying eq. \rf{4u}  then leads to $n_{21}= -{7 \ov 4}$  ( $n_{03}=-\frac{1}{2}$).

Making a natural guess about the structure of the leading term in the 2-loop 
correction to he slope function we then get 
\be 
&&E=E_0+E_1+E_2+... \no \\
&& \ \ \ =  2\sqrt{1+\rrho^{2}}\sqrt\lambda\,\sqrt\mathcal S\Big[
1+\frac{1}{2(\rrho^{2}+1)}\,\mathcal S+\frac{8\rrho^{6}-4\rrho^{4}-16\rrho^{2}-5}{8(\rrho^{2}+1)^{2}}
\,\mathcal S^{2}+...
\Big] \no \\
&&\ \ \ \ \ \ +\frac{n'_{12}-3\zeta_{3}}{\sqrt{\rrho^{2}+1}}\,\mathcal S^{3/2}+... + 
 \frac{1}{\sqrt\lambda}\,\frac{ n_{21}}{(\rrho^{2}+1)^{3/2}}\sqrt\mathcal S+...\ . \la{e2}
\end{eqnarray}


\section*{Appendix D: One-loop energy of the $(J', J)$
 folded string from the algebraic curve approach }

\refstepcounter{section}
\def\theequation{D.\arabic{equation}}
\setcounter{equation}{0}

Here  we shall derive the 1-loop coefficients in \rf{13} 
in  the small spin expansion of the energy of a folded  string with spin $J_1=J'$ and orbital
momentum $J_3=J$  representing a  state in the $su(2)$ sector on the dual gauge theory side. 
This will a  direct counterpart of the computation done for the $(S,J)$ folded string in
 \cite{gva}.

\subsection{Quasimomenta}

The classical solution \cite{Frolov:2003xy} for the 
folded string with spin  $J'$ and orbital momentum  $J$ in $S^5$
is related to the folded string with spin $S$ in $AdS_5$ and  orbital momentum $J$ in $S^5$
by an analytical continuation 
\cite{bfst} implying a relation between  the string profiles
 and the global conserved charges
%
\begin{equation}	
  \left( E ; J', J \right) \to \left( -J; S, -E \right) \ .	
\end{equation}
In the algebraic curve approach the quasimomenta for the 
$(J', J)$ string can then  be obtained by an 
analytical continuation of the quasimomenta for the $(S, J)$ string given in ~\cite{gssv}. 
According to ~\cite{bm}  the $S^5$  quasimomentum
 $p_{\widetilde 2}$  
as a function of the branch points is expressed in terms of the elliptic functions:
\begin{eqnarray}
 p_{\widetilde 2}(x)&=& \pi-
 i\,
{2\pi\E_0}\left(\frac{a}{a^{2}-1}-\frac{x}{x^{2}-1}\right)\,\sqrt{\frac{b}{a}\,\frac{a^{2}-1}{b^{2}-1}}\sqrt{\frac{|a|-i\,a}{|a|-i\,
\overline a}\,\frac{\overline a-x}{a-x}}\,
 \sqrt{\frac{a}{\overline a}\frac{|a|-i\,\overline
a}{|a|-i\,a}\,\frac{\overline a+x}{a+x}}
 \nonumber\\
 && -\frac{8\pi a b \J'}{\,
 (b-a)(a b+1)}\,\mathbf F_{1}(x)-\frac{2\pi
\E_0\,(a-b)}{\,\sqrt{(a^{2}-1)(b^{2}-1)}}\,\mathbf F_{2}(x), \\
 && \phantom \nonumber\\
 {\mathbf F}_{1}(x) &=& i\,\mathbb
F\Big(i\,\sinh^{-1}\sqrt{-\frac{a-b}{a+b}\,\frac{a-x}{a+x}},
  \frac{(a+b)^{2}}{(a-b)^{2}}
 \Big), \\
\mathbf F_{2}(x) &=& i\,\mathbb E\Big(i\,\sinh^{-1}
 \sqrt{-\frac{a-b}{a+b}\,\frac{a-x}{a+x}}, \frac{(a+b)^{2}}{(a-b)^{2}}
\Big),
\end{eqnarray}
where $\mbox{Re}(a), \mbox{Im}(a)>0, \; b = -\overline a$ and 
\begin{eqnarray}
  \J &=& {1\ov 2 \pi}\,\frac{ab-1}{ab}\,\Big[b\,\mathbb E\Big(1-\frac{a^{2}}{b^{2}}\Big)
    +a\,\mathbb K\Big(1-\frac{a^{2}}{b^{2}}\Big)\Big], \nonumber \\
  \J' &=& - {1\ov 2 \pi}\,\frac{ab+1}{ab}\,\Big[b\,\mathbb E\Big(1-\frac{a^{2}}{b^{2}}\Big)
    -a\,\mathbb K\Big(1-\frac{a^{2}}{b^{2}}\Big)\Big], \\
  \E_0 &=& -\frac{1}{\pi b}\,\sqrt{(a^{2}-1)(b^{2}-1)}\,\mathbb K\Big(1-\frac{a^{2}}{b^{2}}\Big).\nonumber
\end{eqnarray}
The inversion symmetry provides the other sphere quasimomenta through the relations 
\begin{equation}
  p_{\widetilde 2}(x) = -p_{\widetilde 3}(x) = -p_{\widetilde 1}({x^{-1}}) = p_{\widetilde 4}({x^{-1}}).
\end{equation} 
Since the motion in the $AdS_5$ part is trivial,
 the corresponding  quasimomenta are  simply 
\begin{equation}
  p_{\widehat 1, \widehat 2}(x) = -p_{\widehat 3, \widehat 4}(x) = 
 2\pi  {\E_0}\,\frac{x}{x^{2}-1}\ . 
\end{equation}


\subsection{Off-shell frequencies}

The symmetry of the solution allows to 
express all the off-shell fluctuation frequencies as 
combinations of only two independent functions~\cite{bm}:
\begin{eqnarray}
  \Omega_{A}(x) &=& \frac{2}{x^{2}-1}\Big(1+x\,\frac{f(1)-f(-1)}{f(1)+f(-1)}\Big), \\
  \Omega_{S}(x) &=& \frac{4}{f(1)+f(-1)}\Big(\frac{f(x)}{x^{2}-1}-1\Big),
\end{eqnarray}
where  $(f(x))^{2} = (x-a)(x-\overline a)(x-b)(x-\overline b)$.
The complete list of the frequencies is given by:
\begin{eqnarray}
  && \Omega^{\widetilde 2\,\widetilde 3}(x) = \Omega_{S}(x),\  \ \ \ \ \ \ \ 
   \Omega^{\widehat 2\,\widehat 3}(x) = \Omega_{A}(x),\nonumber\\
  %
  &&\Omega^{\widetilde 1\, \widetilde 4}(x) = -\Omega_{S}({x^{-1}})+\Omega_{S}(0),\nonumber\\
  &&\Omega^{\widetilde 2\, \widetilde 4}(x) = \Omega^{\widetilde 1\, \widetilde 3}(x) = 
  \frac{1}{2}[\Omega_{S}(x)-\Omega_{S}({x^{-1}})+\Omega_{S}(0)],\nonumber\\
  &&\Omega^{\widehat 1\, \widehat 4}(x) = 
  \Omega^{\widehat 2\, \widehat 4}(x) = \Omega^{\widehat 1\,\widehat 3}(x) = \Omega^{\widehat 2\,\widehat 3}(x),\nonumber\\
  %
  &&\Omega^{\widehat 2\, \widetilde 4}(x) = \Omega^{\widetilde 1\, \widehat 3}(x) = 
  \Omega^{\widetilde 1\, \widehat 4}(x) =\Omega^{\widehat 1\, \widetilde 4}(x) = 
  \frac{1}{2}[\Omega_{A}(x)-\Omega_{S}({x^{-1}})+\Omega_{S}(0)],\nonumber\\
  &&\Omega^{\widetilde 2\, \widehat 4}(x) = \Omega^{\widehat 1\, \widetilde 3}(x) = 
  \Omega^{\widehat 2\, \widetilde 3}(x) = \Omega^{\widetilde 2\, \widehat 3}(x) = 
  \frac{1}{2}[\Omega_{S}(x)+\Omega_{A}(x)].
\end{eqnarray}
The off-shell frequencies provide the fluctuation energies when evaluated on the solutions of the equations:
\begin{equation}
  p_i(x^{i,j}_n) - p_j(x^{i,j}_n) = 2\, \pi\, n. 
\end{equation}

\def \tt {{\rm t}}


\subsection{One-loop correction to the energy}

We have computed the one-loop energy correction $E_{1}$ in the two limits. The first
one  is motivated by 
the analysis in~\cite{bem} and is defined as 
\begin{equation}
\mathcal J'\to 0, \qquad
\qquad \tt\equiv \frac{\mathcal J}{\sqrt{2\,\mathcal J'}} =
\mbox{fixed}\ .
\end{equation}
In this limit, the classical energy is given by 
\begin{equation}
\frac{\mathcal E_{0}}{\sqrt{2\,\mathcal J'}} = 
\sqrt{\tt^2+1}+\frac{ 4 \tt^2+1 }{8 \sqrt{\tt^2+1}}\,\mathcal J'+
\frac{-32\tt^{6}-16\tt^{4}+28\tt^{2}+3}{128\,(\tt^{2}+1)^{3/2}}\,\mathcal J'^{2}+...\ .
\end{equation}
For the one-loop correction we find 
 \be
&& E_{1} = \sum_{p\ge 0}
 a_{p}(\tt)\,(\mathcal J')^{p+\frac{1}{2}} = a_{0}(\tt)\,(\mathcal J')^{1/2}+ 
 a_{1}(\tt)\,(\mathcal J')^{3/2}+... \ , \\
&& a_{0}(\tt) = \displaystyle \frac{1}{2\,\sqrt{2\,(\tt^{2}+1)}}, \ \ \ \ \ \ \ \
a_{1}(\tt) = \displaystyle -\frac{16\,\tt^{4}+25\,\tt^{2}+6}{8\,\big[2\,(\tt^{2}+1)\big]^{3/2}}
-\frac{3}{2\,\sqrt{2\,(\tt^{2}+1)}}\,\zeta_3.
\ee
Adding the classical energy and re-expanding at large $\lambda$ for fixed $J', J$, this gives
\begin{equation}
E^{2} = 2\sqrt\lambda \,J' + \frac{1}{2}{J'^{2}} +J'+J^{2}+\frac{1}{\sqrt\lambda}\Big[
\frac{1}{8}J'^{3} +J'\,J^{2}+\big(
-\frac{5}{8}-3\,\zeta_3
\big)\,J'^{2}+\frac{1}{8}J'
\Big]+... \ , 
\end{equation}
leading to the values of the  coefficients 
$n_{ij}$ in \rf{13}.
The resulting value \be 
n_{12}' = -\frac{5}{8} \ee  is perfectly consistent with the universality
of the two-loop coefficient $n_{21}$ in \rf{9}, i.e. as follows from \rf{4u}, 
\begin{equation}
n_{21} = -\frac{1}{4}\ .
\end{equation}
As  in \cite{bem}, expanding $E_1$ at large $\tt$
we recover the expansion in small $\J'$ for fixed small $\J$:
\begin{eqnarray}
\label{eq:su2-rho}
&&E_{1} =  \Big(\frac{1}{2 {\mathcal J}}-\frac{1}{2}{\mathcal J}+...\Big)\, {\mathcal J'}+
\Big(-\frac{1}{2 {\mathcal J}^3}+\frac{-\frac{1}{8}- {3 \zeta_3} }{2{\mathcal J}}
+...\Big)\, {\mathcal J'}^2 \no \\
&&+ \Big(\frac{3}{4
   {\mathcal J}^5}+\frac{\frac{3}{8}+ {3 \zeta_3}}{2{\mathcal J}^3}+...\Big)\, {\mathcal J'}^3+
    \Big(-\frac{5}{4 {\mathcal J}^7}+\frac{-\frac{23}{8} -  {9 
   \zeta_{3}} }{4{\mathcal J}^5}+...\Big)\, {\mathcal J'}^4+...\ .
\end{eqnarray}
The second limit is 
\begin{equation}
\mathcal J'\to 0\ , \qquad
\qquad \mathcal J = \mbox{fixed} \ .
\end{equation}
In this limit, the classical energy reads\foot{Equivalently, 
$\E^2_0= \J^2  + 2 \sql \sqrt{1 + \J^2} \, \J' + 
{ 1 + 2 \J^2 \ov 2 ( 1 + \J^2)} \, \J'^2 + ...$.
For comparison, in the $(S,J)$ folded string case 
$\E^2_0= \J^2  + 2 \sql \sqrt{1 + \J^2} \, \S + 
{ 3 + 2 \J^2 \ov 2 ( 1 + \J^2)} \, \S^2 + ...$.}
\begin{equation}
\mc E_{0} = \mathcal{J}+
\frac{\sqrt{\mathcal{J}^2+1} }{\mathcal{J}}\,\mc J'-
\frac{3 \mathcal{J}^2+2 }{4\mathcal{J}^3 \big(\mathcal{J}^2+1\big)}\,\mc J'^{2}+\frac{
15 \mathcal{J}^6+33
   \mathcal{J}^4+28 \mathcal{J}^2+8 }{16 \mathcal{J}^5 \big(\mathcal{J}^2+1\big)^{5/2}}\,
   \mc J'^{3}+...\ .
\end{equation}
For the one loop correction we find
\be
E_{1} = e_{1}(\mc J)\,\mc J'+e_{2}(\mc J)\,\mc J'^{2}+e_{3}(\mc J)\,\mc J'^{2}+\dots~,
\ee
and, at order $\mc J'^{2}$,
\begin{equation}
E_{1}= \frac{\mc J'}{2\,\mc J\,(1+\mc J^{2})}+\Big[
\frac{-21 \mathcal J^4-29 \mathcal J^2+1}{16 \mathcal J^3 \big(\mathcal J^2+1\big)^{5/2}}
-\sum_{n=2}^{\infty}\frac{n^2 \big(\mathcal J^2+2 
n^2-1\big)}{\mathcal J^3 \big(n^2-1\big)^2 \big(\mathcal J^2+n^2\big)^{3/2}}
\Big]\,\mc J'^{2}+... \la{yi}
\end{equation}
This expression is very similar to the one for the $(S,J)$ folded string found in \cite{gva}:
\begin{equation}
E_{1}^{(\S,\J)} = 
-\frac{\mc S}{2\,\mc J\,(1+\mc J^{2})}+\Big[
\frac{3 \mathcal{J}^4+11 \mathcal{J}^2+17}{16 \mathcal{J}^3 \big(\mathcal{J}^2+1\big)^{5/2}}
-\sum_{n=2}^{\infty}\frac{n^2 \big(\mathcal{J}^2+2 n^2-1\big)}{\mathcal{J}^3 \big(n^2-1\big)^2 
\big(\mathcal{J}^2+n^2\big)^{3/2}}
\Big]\,\mc S^{2}+...
\end{equation}
The only differences are in 
 the sign of the first  term  (i.e. the  sign of the 
 1-loop term in the ``slope'' function \rf{666})
  and in the coefficients of the contributions of low modes in the second  term.

Extending the calculation to the order $\J'^{3}$ we find the following correction to $E_{1}$
\begin{eqnarray}
e_{3}(\mc J) &=& \frac{150 \mathcal{J}^8+456 \mathcal{J}^6+202 \mathcal{J}^4+8 \mathcal{J}^2-27}{64 \mathcal{J}^5 \left(\mathcal{J}^2+1\right)^4}  \\
&& + \sum_{n=2}^{\infty} \frac{1}{2 \mathcal{J}^5 \left(\mathcal{J}^2+1\right)^{3/2} \left(n^2-1\right)^4 \left(\mathcal{J}^2+n^2\right)^{5/2}}\,\bigg[
\left(8 \mathcal{J}^4+17 \mathcal{J}^2+10\right) n^{10} \nonumber \\
&& + 2 \left(10 \mathcal{J}^6+9 \mathcal{J}^4-13 \mathcal{J}^2-14\right) n^8+2 \left(3 \mathcal{J}^8-19 \mathcal{J}^6-43 \mathcal{J}^4-17 \mathcal{J}^2+7\right) n^6 \nonumber \\
&& -2 \left(6 \mathcal{J}^8+2
   \mathcal{J}^6-13 \mathcal{J}^4-9 \mathcal{J}^2+2\right) n^4-\mathcal{J}^2 \left(2 \left(\mathcal{J}^4+5 \mathcal{J}^2+7\right) \mathcal{J}^2+7\right) n^2\nonumber
\bigg].
\end{eqnarray}
Expanding the coefficients of each 
power of $\J'$ 
  in \rf{yi} in small  $\mc J$ we get explicitly 
  (here $\N=J'$; cf. \rf{hightrJ1J2},\rf{hightrJ1J2x},\rf{hightrJ1J2y})
\begin{eqnarray}
E_{1} &=& \Big(\frac{1}{2 \mathcal J}-\frac{\mathcal J}{2}+\frac{\mathcal J^3}{2}+\dots
\Big)\,\mc J' \no \\
&& +\Big[
-\frac{1}{2 \mathcal J^3}+\frac{1}{\mc J}\Big(-\frac{1}{16} -\frac{3 }{2}\zeta_{3}\Big)
+\mathcal J \Big(-\frac{9}{32} + \frac{3 }{2}\zeta_{3}+\frac{15 }{8}\zeta_{5}\Big)\nonumber \\
&& \qquad\qquad +
\mathcal J^3 \Big(\frac{125}{128} -\frac{  25 }{16}\zeta_{3}-\frac{15 }{8}\zeta_{5}-
\frac{35}{16} \zeta_{7}\Big)+\dots
\Big]\,\mc J'^{2} \nonumber \\
&& +\Big[
\frac{3}{4\,\mc J^{5}}+\frac{1}{\mc J^{3}}\Big(\frac{3}{16} + \frac{3}{2}\,\zeta_{3}
\Big)+
\frac{1}{\mc J}\,\Big(\frac{1}{32}-\frac{9}{8}\,\zeta_{3}\Big)\nonumber \\
&&\qquad\qquad +\mc J\,\Big(\frac{1}{8} + 
3 \zeta_{3}+\frac{35}{16}\, \zeta_{5}-\frac{35}{16}\, \zeta_{7}
\Big)+\cdots
\Big]\,\mc J'^{3}+\dots~. \la{999}
\end{eqnarray}
This is  in perfect agreement with the expansion (\ref{eq:su2-rho})
found  in the  case of  fixed $\tt=\frac{\mc J}{\sqrt{2\,\mc J'}}$. 

From this expansion one extracts,  in particular,  the following values (cf.
\rf{05},\rf{552},\rf{tran})
\be
n_{12} = -\frac{5}{8}-3\,\zeta_{3}\ , 
\qquad 
{\tilde n}_{12} = -\frac{3}{16} + 3 \zeta_3 +\frac{15}{4} \zeta_5\ , 
\qquad
n_{13} = -\frac{7}{16}-\frac{3}{4}\,\zeta_{3}+\frac{15}{4}\,\zeta_{5} \ . \la{jj}
\ee
For comparison, the corresponding values for the $(S,J)$ folded 
string that follow from the analog of \rf{999} in \cite{gva} are:
\be
n_{12} = \frac{3}{8}-3\,\zeta_{3}\ , 
\qquad 
{\tilde n}_{12} = -\frac{27}{16} + 3 \zeta_3 +\frac{15}{4} \zeta_5\ , 
\qquad
n_{13} = -\frac{9}{16}+\frac{15}{4}\,\zeta_{3}+\frac{15}{4}\,\zeta_{5} \ . \la{ss}
\ee
The value of $n''_{13}=-\frac{3}{4}$ in  \rf{uu} for the folded $(J',J)$ string in  \rf{jj} 
is the same as for the $J_{1}=J_{2}$ circular string found in sect 2.2;
 $n''_{13}=\frac{15}{4}$  for the folded $(S,J)$ string in  \rf{ss} 
is the same as for the $S_{1}=S_{2}$ circular string found in sect 2.3.

Similarly  to the  cases of the $(S,J)$ 
folded string \cite{gva} and the circular strings discussed 
in section~2, the coefficient of 
$\J'{}^3/\J$ in \rf{999} 
does not contain $\zeta_{5}$, supporting 
 the universality of the transcendental terms in 
${\tilde n}_{12}$ in \rf{155} and  of the $\zeta_{5}$
term in $n_{13}$ in \rf{uu}. 
Note also  that 
 the highest transcendentality $\zeta_7$ term in the coefficient 
of $\J\,\J'{}^3$ in \rf{999} is also universal, i.e. 
has the same value $(-35/16)$ as in \cite{gva} and  
in all circular string  cases (cf. \rf{hightrJ1J2},\rf{hightrJ1J2x},\rf{hightrJ1J2y}).

\section*{Appendix E:  Summary of coefficients  }

\refstepcounter{section}
\def\theequation{E.\arabic{equation}}
\setcounter{equation}{0}

Here we summarize the  known values of the leading  coefficients in $E^2$ in \rf{2} 
for two  single-spin folded  and three  equal-spin circular  solutions.
We omitted the  values of $ \tn'_{12}, n'_{13}$  for the circular $S=J'$ solution 
that  appear  to be scheme-dependent (see section 2.4).
We  added question marks to the values that  were not computed directly  but are expected 
on the basis of universality of the  Konishi multiplet dimension. 
Let us  recall  the definitions of $n'_{km}, n''_{km}$ as rational coefficients in 
 $n_{12}, \tn_{12}, n_{13}$:
\be
n_{12}=n'_{12}-3\zeta_3\ , 
\qquad\quad
{\tilde n}_{12}={\tilde n}'_{12}+3\zeta_3+\frac{15}{4}\zeta_5\ , 
\qquad\quad
n_{13}=n'_{13}+n''_{13}\zeta_3+\frac{15}{4}\zeta_5\ . \no 
\ee

\newcommand{\tableline}[6]{$#1$ & $#2$ & $#3$ & $#4$ & $#5$ & $#6$ \\}

\begin{table}[H]
\centering \renewcommand{\arraystretch}{1.4}
\begin{tabular}{@{}l|ccccc@{}} \toprule
\tableline{n_{ij}}{(S, J)}{(J', J)}{(J_{1}=J_{2}, J)}{(S_{1}=S_{2}, J)}{(S=J', J)} \midrule
\tableline{n_{01}}
{1}{1}{1}{1}{1}
\tableline{\widetilde n_{01}}
{-\frac{1}{4}}{-\frac{1}{4}}{-\frac{1}{4}}{-\frac{1}{4}}{-\frac{1}{4}}
\tableline{n_{02}}
{\frac{3}{2}}{\frac{1}{2}}{0}{2}{1}
\tableline{\widetilde n_{02}}
{-\frac{1}{2}}{\frac{1}{2}}{1}{-1}{0}
\tableline{n_{03}}
{-\frac{3}{8}}{\frac{1}{8}}{0}{-1}{-\frac{1}{2}}
\tableline{n_{04}}
{\frac{31}{64}}{\frac{1}{64}}{0}{2}{\frac{3}{4}}  \midrule
\tableline{n_{11}}
{-1}{1}{2}{-2}{0}
\tableline{\widetilde n_{11}}
{1}{-1}{-2}{2}{0}
\tableline{\overline n_{11}}  
{-1}{1}{2}{-2}{0}
\tableline{n_{12}'}
{\frac{3}{8}}{-\frac{5}{8}}{-\frac{3}{8}}{\frac{13}{8}}{\frac{5}{8} (?)}
\tableline{\widetilde n_{12}'}    
{-\frac{27}{16}}{-\frac{3}{16}}{-\frac{57}{16}}{-\frac{105}{16}}{-}
\tableline{n_{13}'}     
{-\frac{9}{16}}{-\frac{7}{16}}{-\frac{3}{16}}{-\frac{85}{16}}{-}
\tableline{n_{13}''}
{\frac{15}{4}}{-\frac{3}{4}}{-\frac{3}{4}}{\frac{15}{4}}{\frac{3}{2}}  \midrule
\tableline{n_{21}}
{-\frac{1}{4}}{\ \ \ -\frac{1}{4}(?)}{\ \ \ -\frac{1}{4}(?)}{\ \ \ -\frac{1}{4}(?)}{\ \ \ -\frac{1}{4}(?)}
\bottomrule
\end{tabular}
\caption{Summary of coefficients in eq.~(\ref{2}).}
\end{table}

\newpage


\bigskip


\begin{thebibliography}{30}

\parskip=0.pt


\bibitem{rt1}
  R.~Roiban and A.~A.~Tseytlin,
  ``Quantum strings in \adss: strong-coupling corrections to dimension of
  Konishi operator,''
  JHEP {\bf 0911}, 013 (2009)
  [arXiv:0906.4294].
  
  \bi{tt}
   A.~Tirziu and A.~A.~Tseytlin,
  ``Quantum corrections to energy of short spinning string in $AdS_5$,''
  Phys.\ Rev.\  D {\bf 78}, 066002 (2008)
  [arXiv:0806.4758].
  
  \bi{gssv}
  N.~Gromov, D.~Serban, I.~Shenderovich and D.~Volin,
  ``Quantum folded string and integrability: From finite size effects to Konishi dimension,''
  JHEP {\bf 1108}, 046 (2011)
  [arXiv:1102.1040].

  \bi{rt2}
  R.~Roiban and A.~A.~Tseytlin,
  ``Semiclassical string computation of strong-coupling corrections to dimensions of operators in Konishi multiplet,''
  Nucl.\ Phys.\ B {\bf 848}, 251 (2011)
  [arXiv:1102.1209].

  
  \bi{bm}
  M.~Beccaria and G.~Macorini,
  ``Quantum folded string in $S^5$ and the Konishi multiplet at strong coupling,''
  JHEP {\bf 1110}, 040 (2011)
  [arXiv:1108.3480].

  
  \bi{gva}
  N.~Gromov and S.~Valatka,
  ``Deeper Look into Short Strings,''
  JHEP {\bf 1203}, 058 (2012)
  [arXiv:1109.6305 [hep-th]].
  
    \bi{gkv}
  N.~Gromov, V.~Kazakov and P.~Vieira,
  ``Exact Spectrum of Planar ${\cal N}=4$ Supersymmetric Yang-Mills Theory:
  Konishi Dimension at Any Coupling,''
  Phys.\ Rev.\ Lett.\  {\bf 104}, 211601 (2010)
  [arXiv:0906.4240].
  
  \bi{f}
  S.~Frolov,
  ``Konishi operator at intermediate coupling,''
  J.\ Phys.\ A  {\bf 44}, 065401 (2011)
  [arXiv:1006.5032].
  
  \bi{ff}
   S.~Frolov,
``Scaling dimensions from the mirror TBA,''
  arXiv:1201.2317.



  \bi{bas} B.~Basso,
   ``An exact slope for AdS/CFT,''
  arXiv:1109.3154.

  \bi{p}
  A.~M.~Polyakov,
  ``Gauge fields and space-time,''
  Int.\ J.\ Mod.\ Phys.\ A {\bf 17S1}, 119 (2002)
  [hep-th/0110196].
  
    
  \bibitem{t03} A.~A.~Tseytlin,
  ``On semiclassical approximation and spinning string vertex operators in
  $AdS_5 \times S^5$,''
  Nucl.\ Phys.\  B {\bf 664}, 247 (2003)
  [arXiv:hep-th/0304139].
  Int.\ J.\ Mod.\ Phys.\  A {\bf 25}, 319 (2010)
  [arXiv:0907.3238].
  
  
\bi{ps}
B.C.~Vallilo and L.~Mazzucato,
  ``The Konishi multiplet at strong coupling,''
  JHEP {\bf 1112}, 029 (2011)
  [arXiv:1102.1219].

 
 
 
  \bi{ber}
  M.~Bianchi, J.~F.~Morales and H.~Samtleben,
  ``On stringy AdS(5) x S5 and higher spin holography,''
  JHEP {\bf 0307}, 062 (2003)
  [hep-th/0305052].
  N.~Beisert, M.~Bianchi, J.~F.~Morales and H.~Samtleben,
  ``On the spectrum of AdS / CFT beyond supergravity,''
  JHEP {\bf 0402}, 001 (2004)
  [hep-th/0310292].

  
  
  \bi{rt22}
R.~Roiban, A.~Tirziu and A.~A.~Tseytlin,
  ``Two-loop world-sheet corrections in $AdS_{5}\times S^{5}$ superstring,''
  JHEP {\bf 0707}, 056 (2007)
  [arXiv:0704.3638].



\bi{bkk}
B.~Basso, G.~P.~Korchemsky and J.~Kotanski,
  ``Cusp anomalous dimension in maximally supersymmetric Yang-Mills theory at strong coupling,''
  Phys.\ Rev.\ Lett.\  {\bf 100}, 091601 (2008)
  [arXiv:0708.3933].


  
\bi{eden}
B.~Eden, P.~Heslop, G.~P.~Korchemsky, V.~A.~Smirnov and E.~Sokatchev,
  ``Five-loop Konishi in N=4 SYM,''
  arXiv:1202.5733 [hep-th].




  

  \bi{bd} M.~Beccaria, G.~V.~Dunne, V.~Forini, M.~Pawellek and A.~A.~Tseytlin,
  ``Exact computation of one-loop correction to energy of spinning folded string in 
  \adss,''
  J.\ Phys.\ A {\bf 43}, 165402 (2010)
  [arXiv:1001.4018].
  

\bi{bti}
M.~Beccaria and A.~Tirziu,
  ``On the short string limit of the folded spinning string in \adss,''
  arXiv:0810.4127.



  \bi{bt}
 M.~Beccaria, G.~V.~Dunne, G.~Macorini, A.~Tirziu and A.~A.~Tseytlin,
  ``Exact computation of one-loop correction to energy of pulsating strings in 
  $AdS_5 \times  S^5$,''
  J.\ Phys.\ A  {\bf 44} (2011) 015404
  [arXiv:1009.2318].





\bi{ft2}
S.~Frolov and A.~A.~Tseytlin,
  ``Multi-spin string solutions in \adss,''
  Nucl.\ Phys.\  B {\bf 668}, 77 (2003)
  [arXiv:hep-th/0304255].

\bi{ft3}
S.~Frolov and A.~A.~Tseytlin,
  ``Quantizing three-spin string solution in \adss,''
  JHEP {\bf 0307}, 016 (2003)
  [arXiv:hep-th/0306130].


\bi{art}
G.~Arutyunov, J.~Russo and A.~A.~Tseytlin,
  ``Spinning strings in \adss: New integrable system relations,''
  Phys.\ Rev.\ D {\bf 69}, 086009 (2004)
  [hep-th/0311004].


\bibitem{ptt}
  I.Y.~Park, A.~Tirziu and A.~A.~Tseytlin,
  ``Spinning strings in \adss: One-loop correction to energy in SL(2)
  sector,''
  JHEP {\bf 0503}, 013 (2005)
  [arXiv:hep-th/0501203].


\bibitem{bla} 
  M.~Blau, M.~O'Loughlin, G.~Papadopoulos and A.~A.~Tseytlin,
  ``Solvable models of strings in homogeneous plane wave backgrounds,''
  Nucl.\ Phys.\ B {\bf 673}, 57 (2003)
  [hep-th/0304198].
  
  
\bi{fk1} L.~Freyhult and C.~Kristjansen,
  ``Rational three-spin string duals and non-anomalous finite size effects,''
  JHEP {\bf 0505}, 043 (2005)
  [arXiv:hep-th/0502122].

\bibitem{mik} 
  V.~Mikhaylov,
  ``On the Fermionic Frequencies of Circular Strings,''
  J.\ Phys.\ A  {\bf 43}, 335401 (2010)
  [arXiv:1002.1831].


\bi{mtt1} J.~A.~Minahan, A.~Tirziu and A.~A.~Tseytlin,
  ``1/J corrections to semiclassical AdS/CFT states from quantum
  Landau-Lifshitz model,''
  Nucl.\ Phys.\  B {\bf 735}, 127 (2006)
  [arXiv:hep-th/0509071].


\bi{btz} N.~Beisert, A.~A.~Tseytlin and K.~Zarembo,
  ``Matching quantum strings to quantum spins: One-loop versus finite-size
  corrections,''
  Nucl.\ Phys.\  B {\bf 715}, 190 (2005)
  [arXiv:hep-th/0502173].
 
\bi{gk}  N.~Gromov and V.~Kazakov,
  ``Double scaling and finite size corrections in sl(2) spin chain,''
  Nucl.\ Phys.\  B {\bf 736}, 199 (2006)
  [arXiv:hep-th/0510194].


  


\bi{bfst} N.~Beisert, S.~Frolov, M.~Staudacher, A.~A.~Tseytlin,
  ``Precision spectroscopy of AdS/CFT,''
  JHEP {\bf 0310}, 037 (2003)
  [arXiv:hep-th/0308117].
\bi{grig}
D.~Astolfi, G.~Grignani, T.~Harmark and M.~Orselli,
  ``Finite-size corrections to the rotating string and the winding state,''
  JHEP {\bf 0808}, 099 (2008)
  [arXiv:0804.3301].
  
\bibitem{bem} 
  M.~Beccaria and G.~Macorini,
 ``Resummation of semiclassical short folded string,''
  arXiv:1201.0608.


\bibitem{brt} 
  M.~Beccaria, C.~Ratti and A.~A.~Tseytlin,
  ``Leading quantum correction to energy of 'short' spiky strings,''
  arXiv:1201.5033.

  


\bi{gv2}
 N.~Gromov and P.~Vieira,
  ``The \adss superstring quantum spectrum from
   the algebraic curve,''
  Nucl.\ Phys.\ B {\bf 789}, 175 (2008)
  [hep-th/0703191].


\bibitem{LopezArcos:2012gb} 
  C.~Lopez-Arcos and H.~Nastase,
  arXiv:1203.4777 [hep-th].

\bibitem{Gromov:2008ec} 
  N.~Gromov, S.~Schafer-Nameki and P.~Vieira,
  ``Efficient precision quantization in AdS/CFT,''
  JHEP {\bf 0812}, 013 (2008)
  [arXiv:0807.4752].

\bibitem{Frolov:2003xy}
  S.~Frolov and A.~A.~Tseytlin,
  ``Rotating string solutions: AdS/CFT duality in nonsupersymmetric sectors,''
  Phys.\ Lett.\ B {\bf 570} (2003) 96
  [hep-th/0306143].






\bi{PTT}
I.Y.~Park, A.~Tirziu and A.~A.~Tseytlin,
  ``Semiclassical circular strings in AdS(5) and ``long'' gauge field strength operators,''
  Phys.\ Rev.\ D {\bf 71}, 126008 (2005)
  [hep-th/0505130].







 
\end{thebibliography}
\end{document}